\documentclass[12pt]{article}
\usepackage{amsmath}
\usepackage{graphicx}
\usepackage{enumerate}
\usepackage{natbib}
\usepackage{url} 

\newcommand{\blind}{1}

\usepackage{amsfonts}
\usepackage{amsthm}
\usepackage{amsmath}
\usepackage{amscd}
\usepackage[latin2]{inputenc}
\usepackage{t1enc}
\usepackage[mathscr]{eucal}
\usepackage{indentfirst}
\usepackage{graphicx}
\usepackage{graphics}
\usepackage{pict2e}
\usepackage{epic}
\numberwithin{equation}{section}
\usepackage{epstopdf} 
\usepackage[ruled,vlined]{algorithm2e}
\usepackage{float}

\newcommand \my{\mathrm{y}}
\newcommand \mx{\mathrm{x}}
\newcommand{\ind}{\mathbbm 1}
\newcommand{\V}{\mbox{Var}}
\newcommand{\indi}{\mathbbm I}

\RequirePackage[OT1]{fontenc}
\RequirePackage{amsthm,amsmath}
\RequirePackage[colorlinks,citecolor=blue,urlcolor=blue]{hyperref}
\usepackage{mathrsfs,bbm,bm,amssymb}
\usepackage{ulem}
\usepackage{longtable}
\allowdisplaybreaks
\usepackage{framed}
\usepackage{soul}
\usepackage{enumitem}
\usepackage{tikz}
\usetikzlibrary{shapes,snakes, bayesnet}
\makeatletter
\renewcommand*\env@matrix[1][*\c@MaxMatrixCols c]{%
  \hskip -\arraycolsep
  \let\@ifnextchar\new@ifnextchar
  \array{#1}}
\makeatother

\numberwithin{equation}{section}
\theoremstyle{plain}
\newtheorem{theorem}{Theorem}[section]

\newtheorem{lemma}{Lemma}[section]
\newtheorem{assumption}{Assumption}[section]

\newtheorem{proposition}{Proposition}[section]

\def\ind{\mathbbm{1}}

\newcommand \bbE{\mathbb{E}}

\newcommand \cX{\mathcal{X}}

\usepackage{tikz}
\usetikzlibrary{shapes,snakes, bayesnet}

\graphicspath{{figs/}}

\usepackage{authblk}
\usepackage{xr}

\usepackage{caption,setspace} 
\begin{document}

\def\spacingset#1{\renewcommand{\baselinestretch}%
{#1}\small\normalsize} \spacingset{1}


\if1\blind
{
  \title{\bf Covariate-Assisted Bayesian Graph Learning for Heterogeneous Data}
  \author{Yabo Niu$^\dagger$,
    Yang Ni$^\ddagger$,
    Debdeep Pati$^\ddagger$,
    Bani K. Mallick$^\ddagger$ \\
    $^\dagger$Department of Mathematics, University of Houston \\
    $^\ddagger$Department of Statistics, Texas A\&M University}
    \date{}
  \maketitle
} \fi

\if0\blind
{
  \bigskip
  \bigskip
  \bigskip
  \begin{center}
    {\LARGE\bf Covariate-Assisted Bayesian Graph Learning for Heterogeneous Data}
\end{center}
  \medskip
} \fi

\bigskip
\begin{abstract}
In a traditional Gaussian graphical model, data homogeneity is routinely assumed with no extra variables affecting the conditional independence. In modern genomic datasets, there is an abundance of auxiliary information, which often gets under-utilized in determining the joint dependency structure. In this article, we consider a Bayesian approach to model undirected graphs underlying  heterogeneous multivariate observations with  additional assistance from covariates. Building on product partition models, we propose a novel covariate-dependent Gaussian graphical model that allows graphs to vary with covariates so that observations whose covariates are similar share a similar undirected graph. To efficiently embed Gaussian graphical models into our proposed framework, we explore both Gaussian likelihood and pseudo-likelihood functions. For Gaussian likelihood, a G-Wishart distribution is used as a natural conjugate prior, and for the pseudo-likelihood, a product of Gaussian-conditionals is used. Moreover, the proposed model has large prior support and is flexible to approximate any $\nu$-H\"{o}lder conditional variance-covariance matrices with $\nu\in(0,1]$. We  further show that based on the theory of fractional likelihood,  the rate of posterior contraction is minimax optimal assuming the true density to be a Gaussian mixture with a known number of components. 
The efficacy of the approach is demonstrated via simulation studies and an analysis of a protein network for a breast cancer dataset assisted by mRNA gene expression as covariates.
\end{abstract}

\noindent%
{\it Keywords:} Product partition model; Gaussian graphical model; pseudo-likelihood; G-Wishart prior; posterior contraction rate.
\vfill

\newpage
\spacingset{1.9} 

\section{Introduction}
Graphical models are widely recognized as a powerful tool \citep{dempster1972covariance}  to uncover conditional independence relationship in multivariate observations. They have found widespread applications across many fields, including genomics, causal inference, speech recognition, and computer vision. For instance, in systems biology, graphical models have been used to reverse engineer molecular networks from multi-omic data  \citep{dobra2004sparse, telesca2012modeling, friedman2004inferring}. 

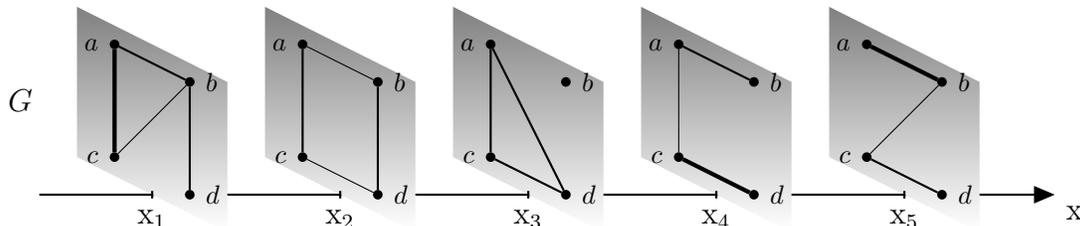
\begin{figure}[H]
  \centering
  \begin{tikzpicture}[thick]

  \draw[thick,->] (-0.5,0.5) -- (13,0.5) node[anchor=north west] {$\mathrm{x}$};
  \shade[black] (0,3) -- (0,1) -- (2,0) -- (2,2) -- cycle;
  \shade[black] (2.5,3) -- (2.5,1) -- (4.5,0) -- (4.5,2) -- cycle;
  \shade[black] (5,3) -- (5,1) -- (7,0) -- (7,2) -- cycle;
  \shade[black] (7.5,3) -- (7.5,1) -- (9.5,0) -- (9.5,2) -- cycle;
  \shade[black] (10,3) -- (10,1) -- (12,0) -- (12,2) -- cycle;

  \node[draw, shape=circle, fill=black, inner sep=1pt, label={left: $a$}] at(0.5,2.5) (g1na) {};
  \node[draw, shape=circle, fill=black, inner sep=1pt, label={left: $c$}] at(0.5,1) (g1nc) {};
  \node[draw, shape=circle, fill=black, inner sep=1pt, label={right: $d$}] at(1.5,0.5) (g1nd) {};
  \node[draw, shape=circle, fill=black, inner sep=1pt, label={right: $b$}] at(1.5,2) (g1nb) {};

  \edge[-, ultra thick] {g1na} {g1nc};
  \edge[-] {g1na} {g1nb};
  \edge[-, thin] {g1nc} {g1nb};
  \edge[-] {g1nb} {g1nd};

  \node[draw, shape=circle, fill=black, inner sep=1pt, label={left: $a$}] at(3,2.5) (g2na) {};
  \node[draw, shape=circle, fill=black, inner sep=1pt, label={left: $c$}] at(3,1) (g2nc) {};
  \node[draw, shape=circle, fill=black, inner sep=1pt, label={right: $d$}] at(4,0.5) (g2nd) {};
  \node[draw, shape=circle, fill=black, inner sep=1pt, label={right: $b$}] at(4,2) (g2nb) {};

  \edge[-, thin] {g2na} {g2nb};
  \edge[-] {g2na} {g2nc};
  \edge[-, thin] {g2nc} {g2nd};
  \edge[-] {g2nb} {g2nd};

  \node[draw, shape=circle, fill=black, inner sep=1pt, label={left: $a$}] at(5.5,2.5) (g3na) {};
  \node[draw, shape=circle, fill=black, inner sep=1pt, label={left: $c$}] at(5.5,1) (g3nc) {};
  \node[draw, shape=circle, fill=black, inner sep=1pt, label={right: $d$}] at(6.5,0.5) (g3nd) {};
  \node[draw, shape=circle, fill=black, inner sep=1pt, label={right: $b$}] at(6.5,2) (g3nb) {};

  \edge[-] {g3na} {g3nd};
  \edge[-] {g3na} {g3nc};
  \edge[-] {g3nc} {g3nd};

  \node[draw, shape=circle, fill=black, inner sep=1pt, label={left: $a$}] at(8,2.5) (g4na) {};
  \node[draw, shape=circle, fill=black, inner sep=1pt, label={left: $c$}] at(8,1) (g4nc) {};
  \node[draw, shape=circle, fill=black, inner sep=1pt, label={right: $d$}] at(9,0.5) (g4nd) {};
  \node[draw, shape=circle, fill=black, inner sep=1pt, label={right: $b$}] at(9,2) (g4nb) {};

  \edge[-] {g4na} {g4nb};
  \edge[-, ultra thick] {g4nc} {g4nd};
  \edge[-, thin] {g4nc} {g4na};

  \node[draw, shape=circle, fill=black, inner sep=1pt, label={left: $a$}] at(10.5,2.5) (g5na) {};
  \node[draw, shape=circle, fill=black, inner sep=1pt, label={left: $c$}] at(10.5,1) (g5nc) {};
  \node[draw, shape=circle, fill=black, inner sep=1pt, label={right: $d$}] at(11.5,0.5) (g5nd) {};
  \node[draw, shape=circle, fill=black, inner sep=1pt, label={right: $b$}] at(11.5,2) (g5nb) {};

  \edge[-, ultra thick] {g5na} {g5nb};
  \edge[-] {g5nc} {g5nd};
  \edge[-, thin] {g5nc} {g5nb};

  \draw (1,0.55) -- (1,0.45) node[anchor=north] {$\mathrm{x}_1$};
  \draw (3.5,0.55) -- (3.5,0.45) node[anchor=north] {$\mathrm{x}_2$};
  \draw (6,0.55) -- (6,0.45) node[anchor=north] {$\mathrm{x}_3$};
  \draw (8.5,0.55) -- (8.5,0.45) node[anchor=north] {$\mathrm{x}_4$};
  \draw (11,0.55) -- (11,0.45) node[anchor=north] {$\mathrm{x}_5$};
  \node[] at(-0.75, 1.75) {$G$};

  \end{tikzpicture}
  \vspace{-5mm}
  \caption{\small \relscale{0.9} A toy example of an undirected graphical model with four nodes ($a$, $b$, $c$, and $d$) and one continuous covariate $\mathrm{x}$. The edge thickness is proportional to its strength. Both  the  structure  and  the  edge  strength vary with the covariate. } \label{fig:toy}
\end{figure}

Typically, a graphical model assumes a graph-dependent probability distribution $\mathcal{P}_G(\mathrm{y})$ as the sampling distribution for a $q$-dimensional random vector $\mathrm{y}=(y_1, \ldots, y_q)$, e.g., a centered Gaussian distribution with a sparse precision or inverse-covariance matrix. We assume that the conditional independence relationships in $\mathrm{y}$ are encoded via an undirected graph $G=(V, E)$ that consists of a set of nodes $V=(1,\ldots, q)$ and a set of undirected edges $E$ such that $s-t\not\in E$ if $y_s$ and $y_t$ are conditionally independent given all other variables.
One common assumption of many existing graphical model approaches is that the observations are homogeneous and a single graph $G$ is sufficient to characterize the conditional dependency structure of $\mathrm{y}$. However, such an assumption can be violated in many modern applications such as cancer genomic studies where patients are known to be highly heterogeneous and may have distinct molecular networks whose structures are modulated by various individualized factors such as genetic markers and prognostic factors \citep{lohr2014widespread, bolli2014heterogeneity, dahl2008distance}. 
Existing statistical approaches (briefly reviewed below) do not adequately capture such network plasticity for heterogeneous populations and hence the goal of this article is to fill in the gap by developing novel covariate-dependent undirected graphical models.
We illustrate the idea of the proposed model with a toy example  in  \figurename{ \ref{fig:toy}} where a 4-node undirected graph changes with covariate $\mathrm{x}$ in both its structure and edge strength. 
\vspace{-5mm}
\noindent \textbf{Existing literature on heterogeneous graphical models.}
\citet{guo2011joint,danaher2014joint,peterson2015bayesian,ha2015dingo} considered heterogeneous graphical models without covariates to estimate multiple group-specific undirected graphs where the groups are pre-fixed. The approaches depend on the specific criteria used to segment heterogeneous data, which are often subjective and not necessarily always available, and may result in groups that are still heterogeneous.
In another line of work, \citet{rodriguez2011sparse,telluri} explored mixtures of graphical models  where the population is objectively clustered into homogeneous groups with each group having different graph structures. The clustering is mainly driven by the separation of group-specific means instead of graphs. As demonstrated in our simulation studies, covariate information can be beneficial to the estimation of multiple graphs even when there is no clear separation in the first moment of the data.  Refer also to \cite{yin2011sparse, lee2012simultaneous,bhadra2013joint, cai2013covariate,ni2018reciprocal, deshpande2019simultaneous, niu2020bayesian, samanta2022generalized} for graphical models with covariate-dependent mean structure. 
In essence, these approaches can be written as a multivariate regression model with residuals following a graphical model. 
While they are effective in adjusting covariates for graph estimation, the graph structure is still assumed to be homogeneous across all observations. 

As an extension from covariate dependency through the first moment, a second order covariance regression framework 
\citep{hoff2012covariance,fox2015bayesian} 
models the covariance matrix $\Sigma_x$ as a function of covariates, $\Sigma_x=\Lambda_x\Lambda_x^T+\Psi$ for some positive diagonal matrix $\Psi$. 
Matrix $\Sigma_x$ is clearly positive-definite at any value of $x$ by construction. Although, in principle, the covariance regression framework can be extended to graph or inverse-covariance-regression by assuming $\Omega_x=\Sigma_x^{-1}=\Lambda_x\Lambda_x^T+\Psi$, it is not immediately clear how to induce sparsity in $\Omega_x$, a key feature of undirected Gaussian graphical models (GGMs). For instance, in order to set the $(i,j)$th element of $\Omega_x$ to zero, $\Lambda_x$ needs to be carefully chosen so that the inner product of the $i$th and $j$th rows of $\Lambda_x$ is exactly zero. Such systematic control calls for a tedious modeling exercise, which becomes even more challenging in relatively sparse high dimensional $\Omega_x$.


There has been some recent developments for covariate-dependent graphs and sparse precision matrices. \citet{liu2010graph} developed a graph-valued regression model that partitions the covariate space into rectangles by classification and regression trees and fits GGMs separately to each region. However, the estimated graphs may become unstable and lack similarity for similar covariates due to the separate graph estimation, as reported in \cite{Cheng14}. Several kernel-based methods \citep{kolar2010sparse,kolar2010estimating,zhou2010time} have been developed for conditional precision matrix estimation. However, due to the curse of dimensionality, they all focused on a univariate covariate. Recently,  \citet{ni2019bayesian} proposed a graphical regression method that estimates directed acyclic graphs as functions of covariates and allows the graph structure to vary continuously with covariates. However, it is difficult to extend their approach to undirected graphs because as mentioned earlier, formulating a sparse positive-definite matrix directly as a function of covariates is highly nontrivial. To the best of our knowledge, there is no existing method that can model continuously varying undirected graphs as functions of general covariates.

\noindent \textbf{Outline of the proposed approach.} 
We propose a novel Bayesian covariate-dependent graphical model that circumvents the challenges described above. Specifically, we avoid directly parameterizing the graphs or sparse precision matrices as functions of covariates by introducing an intermediate layer of latent variables and exploiting the conditional independence of graph and covariates given the intermediate latent variables. We choose this layer to be a random partition that serves as a hidden link between the graphs and covariates. We then specify a model for the graphs given the random partition and a model for the random partition given the covariates, both of which are significantly simpler than directly modeling sparse precision matrices as functions of covariates. Despite the discrete nature of the random partition, one can still arrive at a continuously varying graph or precision matrix by marginalizing out the covariate-dependent partition. We take note here that this cannot be done in non-probabilistic partition-based approaches such as \cite{liu2010graph}.  Even in applications where a clear partition of the population is deemed plausible, marginalizing out the partition can still be interesting and provides natural uncertainty quantification of the graph estimation assisted by covariates. While our focus in this paper is on undirected graphs, the proposed model can be easily extended to other graphs such as directed acyclic graphs. 

Moreover,  we study the theoretical support of our proposed prior by showing that our model is flexible to approximate any $\nu$-H\"{o}lder conditional variance-covariance functions $\nu\in(0,1]$. It is well-known that such functions are dense (in an $L_1$ sense) in the class of all continuous functions, attesting to the large support.  
 We also establish optimal rates of fractional posterior contraction by using a specific choice of the cohesion function that builds connection between the conditional and joint densities, which allows us to leverage on the existing results on posterior convergence rates of mixture models.

\noindent 
\textbf{Novelties of the proposed approach.} 
\textbf{(1) Structurally varying graphs.} The existing methods can only handle the covariate-dependent undirected graphs in a discrete fashion \citep{liu2010graph}. 
Our approach offers the ability to model a continuously varying graph by marginalizing out the discrete partition so that discretely and continuously varying graphs are unified seamlessly under one Bayesian framework, which has not been achieved before.
\textbf{(2) Theoretical results.} The large support result of the proposed prior shows that although the prior for the conditional precision matrix is continuous in covariates through the multinomial logit transformation in probabilities, it assigns positive probabilities to any $L_1$ neighborhood of a  $\nu$-H\"{o}lder continuous conditional precision matrix ($0<\nu\leq 1$, see its definition in \S \ref{priorsupport}). To the best of our knowledge, this is the first result on the flexibility of covariate-dependent graphical models on the space of piecewise conditional covariance functions. In addition, our posterior contraction result demonstrates that our posterior for the conditional density is rate-optimal.
\textbf{(3) Generality.} Our approach is a general class of models that includes some of the methods mentioned above as special cases. For example, by using certain cohesion function, \cite{rodriguez2011sparse} can be considered as a special case of our approach without any covariates.
\textbf{(4) Generalizability.} Some methods \citep{liu2010graph} can be adapted to different types of graphs but not without major modifications, and some methods \citep{ni2019bayesian} cannot easily move beyond its specification. Our approach offers an easy adaptation to any kind of graphs without changing anything significant and it can be easily done by replacing the graph-related likelihood only.


The rest of this article is organized as follows. First, we introduce and formulate the general problem of interest in \S \ref{sec:general}. We present our proposed model in \S \ref{sec:PxG}, which contains prior and likelihood specification, model averaging and prediction, and theoretical results of the prior support and the fractional posterior consistency. Posterior inference including a fast blocked Gibbs sampler and some details of model comparison are defered to \S \ref{sec:algo}. In \S \ref{sec:simu}-\S \ref{sec:example}, we illustrate the proposed method using simulations with different dimensions of covariates and graphs, and an application of protein networks in breast cancer with gene expressions as covariates. We conclude this paper with discussions in \S \ref{sec:discussion}.

\section{General Formulation} \label{sec:general}
Let $\mathrm{y}_1,\ldots,\mathrm{y}_n$ be $n$ realizations of a $q$-dimensional random vector $\mathrm{y}=(y_1, \ldots, y_q)$ of primary interest and let $\mathrm{x}_1,\ldots,\mathrm{x}_n$ be the realizations of $p$-dimensional secondary covariates $\mathrm{x}=(x_1,\dots,x_p)$. The goal is to 
build a graphical model that characterizes the conditional independencies of $\mathrm{y}$ conditional on $\mathrm{x}$. Specifically, we assume 
a covariate-dependent graphical model $\mathrm{y}\sim \mathcal{P}\{\mathrm{y}\mid \Omega(\mathrm{x})\}$
where $\Omega(\mathrm{x})$ encodes the covariate-dependent graph structure. 
For GGMs,  $\Omega(\mathrm{x})$ is the covariate-dependent precision matrix. As mentioned in the introduction, directly modeling a covariate-dependent precision matrix can be a challenging problem: it is difficult to ensure that the precision matrix is both sparse and positive-definite while being smoothly varying with the covariates.

\begin{figure}[h]
  \centering
  \scalebox{0.8}{
  \begin{tikzpicture}[thick]
  \node[draw, circle] (X) {$X$};
  \node[draw, circle, below=2 of X] (T) {$\Omega(\mathrm{x})$};
  \edge[->] {X} {T};
  \node[draw, circle, below=0.75 of T] (Y) {$Y$};
  \edge[->] {T} {Y};
  \node[left=0.5 of X] {covariates};
  \node[left=0.5 of T] {parameters};
  \node[left=0.5 of Y] { $\mathcal{P}\{\mathrm{y}\mid \Omega(\mathrm{x})\}$ };
  \node[below=0.5 of Y] {(a)};

  \node[draw, circle, right=6 of X] (X2) {$X$};
  \node[draw, circle, below=0.5 of X2] (P) {$\bm{\rho}_\mathrm{x}$};
  \node[draw, circle] at (5,-3) (T1) {$\Omega_1$};
  \node[draw, circle, right=0.7 of T1] (T2) {$\Omega_2$};
  \node[draw, circle, right=0.7 of T2] (T3) {$\Omega_3$};
  \node[right=0.5 of T3] (T4) {... ... ...};
  \node[right=0.5 of T4] (T5) {};
  \edge[->] {P} {T1};
  \edge[->] {P} {T2};
  \edge[->] {P} {T3};
  \edge[->] {X2} {P};
  \node[draw, circle, below=0.75 of T2] (Y2) {$Y$};
  \edge[->] {T1} {Y2};
  \edge[->] {T2} {Y2};
  \edge[->] {T3} {Y2};
  \node[below=0.5 of Y2] {(b)};
  \node[right=0.25 of P] {covariate-dependent partition};
  \edge[->] {P} {T4};
  \edge[->] {P} {T5};
  \edge[->] {T4} {Y2};
  \edge[->] {T5} {Y2};

  \node[] at (4,-3) (R) {};
  \node[] at (1,-3) (L) {};
  \edge[->] {R} {L};
  \node[] at (2.7, -2.4) {marginalizing out $\bm{\rho}_\mathrm{x}$};
  \end{tikzpicture}
  }
  \vspace{-0.45cm}
  \caption{\small \relscale{0.9} (a) A direct model. (b) The proposed model with an extra layer.} \label{add2}
\end{figure}
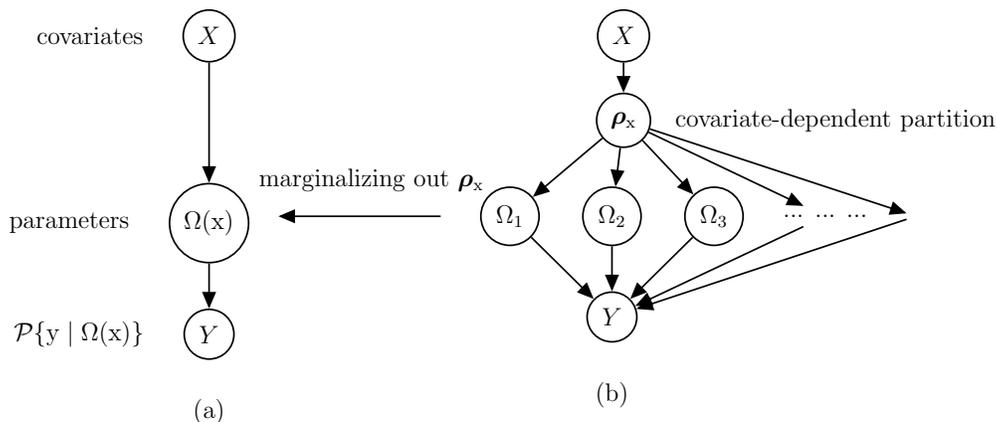
\vspace{-0.5cm}

To address this challenge, we introduce a discrete latent variable -- a covariate-dependent partition -- to avoid explicitly define the precision matrix as a function of covariates. It serves as an additional layer between the precision matrix and covariates. This is natural in a Bayesian hierarchical model, and it can take advantage of marginalization as we shall show in the next section. The covariate-dependent partition ensures the precision matrix function to have only a finite number of values according to the partition; thus, a finite number of graphs as well. To maintain positive definiteness and sparsity in a finite number of precision matrix is substantially easier and many models have been developed for this~purpose. 

Let $\bm{\rho}=(S_1,\ldots, S_{K_{\bm\rho}})$ denote a random partition of $[n]=\{1,\dots,n\}$, i.e., $S_j\cap S_{j'}=\emptyset$ for $j\neq j'$ and $\cup_{j=1}^{K_{\bm\rho}} S_j=[n]$, where $K_{\bm\rho}$ is the size of partition $\bm\rho$. In other words, the set $S_j$ contains the indices of the observations that belong to cluster $j$ and $K_{\bm\rho}$ is the number of clusters. For each cluster $j=1,2,\dots,K_{\bm\rho}$, we let $\Omega_j$ denote its cluster-specific precision matrix and $\bm{\Omega}=\{\Omega_1,\ldots,\Omega_{K_{\bm\rho}}\}$.
The probability distribution of $\mathrm{y}_1,\ldots,\mathrm{y}_n$ given the partition and cluster-specific precision matrices can be written as 
\small
\vspace{-0.7cm}
\begin{equation} \label{eq:ylikelihood1}
  \mathcal{P}(\mathrm{y}_1,\ldots,\mathrm{y}_n \mid \bm{\rho}, \bm{\Omega}) =\prod_{j=1}^{K_{\bm\rho}}\prod_{i\in S_j} \mathcal{P}(\mathrm{y}_i\mid\Omega_j).
\end{equation}
\vskip -0.6cm
\normalfont
\noindent
Notice, by introducing the cluster specific indicators $\bm{z}=\{z_1,\ldots,z_n\}$, where $z_i=j$ if $i\in S_j$ for $i=1,\ldots,n$ and $j=1,\ldots,K_{\bm\rho}$, an equivalent representation of (\ref{eq:ylikelihood1}) can be obtained as
\vskip -0.8cm
\small
\begin{equation*}
    \mathcal{P}(\mathrm{y}_1,\ldots,\mathrm{y}_n \mid \bm{z}, \bm{\Omega}) =\prod_{i=1}^{n} \mathcal{P}(\mathrm{y}_i\mid z_i, \bm{\Omega}) = \prod_{i=1}^{n}\mathcal{P}(\mathrm{y}_i\mid \Omega_{z_i}).
\end{equation*}
\normalfont
\vskip -0.3cm
\noindent
Hence, $\mathrm{y}_i$'s are independent given $\bm z$ and $\bm{\Omega}$, i.e., $\mathrm{y_i}\mid z_i,\bm{\Omega}\sim \mathcal{P}(\mathrm{y}_i\mid \Omega_{z_i})$. Therefore, the true data-generating distribution of $\mathrm{y}$ can be written as a mixture distribution,
\vskip -0.8cm
\small
\begin{equation} \label{eq:ylikelihood2}
    \mathcal{P}(\mathrm{y}\mid \bm{\pi},\bm{\Omega}) = \sum_{j=1}^{K_{\bm\rho}}\pi_j\mathcal{P}(\mathrm{y}\mid\Omega_j),
\end{equation}
\normalfont
\vskip -0.5cm
\noindent
where $\pi_j=P(z_i=j)$ for $i=1,\ldots,n$ and $j=1,\ldots,K_{\bm\rho}$ and $\sum_{j=1}^{K_{\bm\rho}}\pi_j=1$, $\bm{\pi}=\{\pi_1,\ldots,\pi_{K_{\bm\rho}}\}$.

For the choice of graphical models (i.e., the choices of probability distribution $\mathcal{P}(\mathrm{y}_i\mid\Omega_j)$ and the graphical prior $p(\Omega_j)$), we focus on undirected GGMs. We adopt two approaches in this article. First, we consider a Gaussian likelihood function with the G-Wishart prior \citep{roverato2002hyper} as the natural conjugate prior for the precision matrix. It is a well-studied model with many existing efficient procedures \citep{atay2005monte,jones2005experiments, mohammadi2015bayesian,lenkoski2011computational,dobra2011bayesian,wang2012efficient,lenkoski2013direct,uhler2018exact}. However, this natural Bayesian formulation does not scale well with high-dimensional graphs. Traditionally, to scale up, researchers often restrict graphs to be decomposable \citep{dawid1993hyper}; see also some recent theoretical developments \citep{lee2021bayesian,xiang2015high}. Another scalable approach, which does not assume decomposability, is the pseudo-likelihood implementation of the neighborhood selection method for GGMs \citep{meinshausen2006high}. Several Bayesian adaptations have been proposed in recent years \citep{atchade2015scalable, lin2017joint}. The pseudo-likelihood approach, in addition to scalability, allows the use of any existing regression variable selection methods such as the spike-and-slab prior \citep{mitchell1988bayesian, ishwaran2005spike, narisetty2014bayesian}, which we adopt in this article.

To incorporate covariates $\mathrm{x}_1,\dots,\mathrm{x}_n$, we consider covariate-dependent random partition prior models $p(\bm{\rho} \mid \mathrm{x}_1,\ldots, \mathrm{x}_n)$. In this paper we choose to use the product partition model with covariates (PPMx, \citet{muller2011product}). But several other Bayesian  models that incorporate covariate information can be used as well, such as dependent Dirichlet process (DDP) models \citep{maceachern1999dependent, sethuraman1994constructive, griffin2006order, muller1996bayesian, barcella2017comparative} and the hierarchical mixture of experts (HME) models \citep{dasgupta1998detecting, jordan1994hierarchical, bishop2012bayesian}. Neither DDP or HME is equivalent to PPMx, and none of them has been extended for graphical models. To the best of our knowledge, our proposed model is the first Bayesian partition-based model adapted for covariate-dependent graphical models with heterogeneous data. 

Exploiting marginalization, the covariate-dependent graphical model is defined as,
\small
\vspace{-0.8cm}
$$
\mathcal{P}\{\mathrm{y}_1,\dots,\mathrm{y}_n\mid\Omega(\cdot),\mathrm{x}_1\dots,\mathrm{x}_n\}=\sum_{\bm{\rho}\in \mathcal{B}_n} \mathcal{P}(\mathrm{y}_1,\ldots,\mathrm{y}_n \mid \bm{\rho}, \Omega_1,\dots \Omega_{K_{\bm\rho}})p(\bm{\rho}\mid\mathrm{x}_1,\dots,\mathrm{x}_n),
$$
\vskip -0.6cm
\normalfont
\noindent where $\mathcal{B}_n$ is the collection of all possible partitions of $[n]$. It is clear from the marginal likelihood that the precision matrix $\Omega(\cdot)$ is a function of covariates.

\section{Partition-Based Covariate-Dependent Graphical Model} \label{sec:PxG}
The proposed partition-based covariate-dependent graphical model (PxG) is a hierarchical model with three levels (see Figure \ref{add2}): the probability distribution $\mathcal{P}(\mathrm{y}\mid\Omega)$, the graphical prior $p(\Omega\mid\bm{\rho})$ given the partition $\bm{\rho}$, and the covariate-dependent partition prior $p(\bm{\rho}\mid\mathrm{x})$. We will describe the modeling details of each level.

\subsection{Covariate-dependent partition prior} \label{partitionprior}


As stated in \S \ref{sec:general}, we consider partition-based prior models to incorporate covariates. There is a variety of them, such as PPMx, DDP, and HME, just to name a few. In this paper, we consider PPMx, i.e., the product partition model with covariates proposed in \cite{muller2011product} for its simplicity and ease to use. See \citet{page2021discovering} for another application of PPMx. But this does not limit the choice to only PPMx. In fact, any other partition-based models can be adopted within the same Bayesian framework directly if one prefers, which is a big advantage of the proposed model. 
The PPMx model \citep{muller2011product} is defined as
\small
\vspace{-0.6cm}
$$
p(\bm{\rho} \mid \mathrm{x}_1,\ldots, \mathrm{x}_n) \propto \prod_{j=1}^{K_{\bm\rho}} g(\mathrm{X}^*_j)\cdot c(S_j),
$$
\vskip -0.6cm
\normalfont
\noindent where the cohesion function $c(S) \geq 0$ for $S \subseteq [n]$ measures the tightness of the elements in $S$ and the similarity function $g(\mathrm{X}^*_j)\geq 0$ for $\mathrm{X}^*_j=\{\mathrm{x}_i\mid i\in S_j\}$ characterizes the similarity of the $\mathrm{x}_i$'s in cluster $j$.
The similarity function need not be a proper probability model of covariates, but for convenience, we construct it by marginalizing out the parameters in the auxiliary probability model of covariates $p(\mathrm{x}_i\,|\,\Gamma_j)$ within each group, where $\Gamma_j\sim p(\Gamma_j)$ is a set of parameters, hence
\small
\vspace{-0.7cm}
\begin{equation*}
  g(\mathrm{X}^*_j) = \int \prod_{i\in S_j} p(\mathrm{x}_i \mid \Gamma_j ) p(\Gamma_j) d\Gamma_j.
\end{equation*}
\vskip -0.6cm
\normalfont
\noindent Let $z_1, \ldots, z_n\in\{1,\ldots,K_{\bm\rho}\}$ denote the cluster indicator such that $z_i=j$ if $i\in S_j$. The covariates in our later application are continuous, and therefore we use the following default auxiliary probability model of covariates for cluster $j=1,\dots,K_{\bm\rho}$,
\small
\vspace{-0.6cm}
\begin{equation} \label{eq:covf}
\begin{aligned}
  \{\mathrm{x}_i\}_{z_i=j} \mid \bm{\mu}_{j}, \sigma^2_{j} & \stackrel{i.i.d.}{\sim} \prod_{i\in S_j}N_p(\bm{\mu}_{j}, \sigma^2_{j}I_p), \\
  \bm{\mu}_{j} \mid \sigma^2_{j} & \sim N_p(\bm{\mu}_{0}, \sigma^2_{0}\sigma^2_{j}I_p), \\
  \sigma^2_{j} & \sim IG(b_{1}, b_{2}).
\end{aligned}
\end{equation}
\vskip -0.6cm
\normalfont
\noindent The similarity function can be constructed in the same fashion for discrete covariates. Through similarity functions, the PxG model encourages that similar covariates lead to similar clustering, which in turn leads to similar graph structures. The choices of cohesion function $c(\cdot)$ are plenty. For simplicity, we use $c(S_j)=\alpha(n_j-1)!$ where $n_j=|S_j|$ is the size of cluster $j$. With this choice of cohesion function, Dirichlet process (DP) mixture model is a special case of the PxG model \citep{ferguson1973bayesian, antoniak1974mixtures}, and \cite{rodriguez2011sparse} is equivalent to the PxG model with no covariates. See \S \ref{sec:algo} for details of the blocked Gibbs sampler for the PxG model.


\subsection{Probability distribution and graphical prior} \label{sec:likelihood}

We focus on undirected GGMs although our methodology is applicable to a much larger class of graphical models. 
Specifically, conditional on the partition $\bm{\rho}$, we independently assign an undirected graph and precision matrix to each cluster. That is, $\Omega(\mathrm{x}_i)=\Omega_j$ if $i\in S_j$. We effectively reduce an infinite-dimensional functional $\Omega(\cdot)$ to a finite set of parameters $\Omega_1,\dots, \Omega_{K_{\bm\rho}}$.
Let $\mathrm{Y}^*_j = (\mathrm{y}^T_{i_1},\ldots,\mathrm{y}^T_{i_{n_j}})$ with $n_j=|S_j|$ and $S_j=\{i_1,\ldots,i_{n_j}\}$.
We consider two approaches for GGMs, which correspond to two different choices of the probability distribution $\mathcal{P}(\mathrm{Y}^*_j\mid\Omega_j)$ and graphical prior $p(\Omega_j,G_j)$ for $j=1,\dots,K_{\bm\rho}$. 

\subsubsection{Approach one: Gaussian-G-Wishart model}\label{ggw}
The probability distribution is a centered multivariate Gaussian distribution,  $\mathcal{P}(\mathrm{Y}^*_j\mid\Omega_j)=\prod_{i\in S_j}N_q(\mathrm{y}_i\mid 0,\Omega_j^{-1})$ where the precision matrix  $\Omega_j=\{\omega_{st}^j\}_{s,t=1}^q$ encodes the graph structure of $G_j=(V,E_j)$ such that $\omega_{st}^j\neq0$ if and only if $s-t \in E_j$.
Conditional on $G_j$, we assume a conjugate G-Wishart prior $\Omega_j\sim \text{G-Wishart}_{G_j}(b, D)$ with density,
\small
\vspace{-0.7cm}
\begin{equation*}
  p(\Omega_j \mid G_j) = I_{G_j}^{-1}(b, D) |D|^{(b-2)/2}\exp\Big\{-\frac{1}{2}\mbox{tr}(\Omega_j^{-1} D)\Big\},
\end{equation*}
\vskip -0.7cm
\normalfont
\noindent where $I_{G_j}^{-1}(b, D)$ is the analytically intractable normalizing constant and $\mbox{tr}(\cdot)$ is the trace of a matrix. The graphical prior is completed with a prior distribution on graph $G$. Let $g_{st}^j$ be a binary indicator variable such that $g_{st}^j=1$ if $s-t\in E_j$ and 0 otherwise. We assume $g_{st}^j$'s are independent Bernoulli, $g_{st}^j\sim \text{Bernoulli}(\alpha_G)$, for $1\leq s<t \leq q$, with prior edge inclusion probability $\alpha_G$.

Many sampling procedures have been developed for Gaussian-G-Wishart model \citep{jones2005experiments, mohammadi2015bayesian, dobra2011bayesian}. To improve the scalability of posterior inference, we also consider a pseudo-likelihood approach.


\subsubsection{Approach two: pseudo-likelihood model} 
As the graph size increases, the Gaussian-G-Wishart approach becomes computationally expensive. Instead of using proper Gaussian likelihood function that constrains the precision matrix to be positive definite, we replace it by the pseudo-likelihood function that is a product of conditional likelihood functions \citep{besag1975statistical}. Let $\mathrm{y}_{i,-s}$ be a subvector of $\mathrm{y}_i$ without the $s$th element. Denote the conditional likelihood of $\mathrm{y}_{i,s}\mid \mathrm{y}_{i,-s}$ as $Q(\mathrm{y}_{i,s}\mid \mathrm{y}_{i,-s}, \Omega_j)$ for $i\in S_j$, where $\mathrm{y}_{i,s}$ is the $s$th element of $\mathrm{y}_i$. Then the pseudo-likelihood function is defined as
\small
\vspace{-0.5cm}
\begin{equation*}
\mathcal{P}(\mathrm{Y}^*_j\mid\Omega_j)=\prod_{i\in S_j} \prod_{s=1}^{q} Q(\mathrm{y}_{i,s}\mid \mathrm{y}_{i,-s},\Omega_j).
\end{equation*}
\vskip -0.7cm
\normalfont
\noindent Such pseudo-likelihood approach has been adopted in modeling undirected graphical models \citep{ji1996consistent, csiszar2006consistent,pensar2017marginal, barber2015high, ravikumar2010high, ekeberg2013improved}.
For GGMs, the conditional likelihood function is the node-wise linear regression model, i.e., $Q(\mathrm{y}_{i,s}\mid \mathrm{y}_{i,-s},\Omega_j)\equiv N(\mathrm{y}_{i,s}\mid \mathrm{y}_{i,-s}^T\bm{\beta}_s^j, \tau^j_s)$ where $\bm{\beta}_s^j=(\beta^j_{s1},\dots,\beta^j_{s,s-1},\beta^j_{s,s+1},\dots,\beta^j_{sq})^T$.  This choice is motivated by the fact that the conditional density of $N(\mathrm{y}_i\mid 0,\Omega_j^{-1})$ is $N(\mathrm{y}_{i,s}\mid \mathrm{y}_{i,-s}^T\bm{\beta}_s^j, \tau^j_s)$ with $\beta_{st}^j=-\omega_{st}^j/\omega_{ss}^j$ and $\tau_s^j=1/\omega_{ss}^j$ \citep{peng2009partial}. Therefore, $\omega_{st}^j=0$ if and only if $\beta_{st}^j=0$. Consequently, determining the graph structure is equivalent to finding a sparse subset of $\bm{\beta}_s^j$ in the regression model $N(\mathrm{y}_{i,s}\mid \mathrm{y}_{i,-s}^T\bm{\beta}_s^j, \tau^j_s)$.

In essence, the pseudo-likelihood approach converts the graph structure learning task for GGMs into a variable selection problem in a set of independent linear regressions. Thus, any existing variable selection procedures can be adopted here, including both spike-and-slab priors \citep{mitchell1988bayesian, ishwaran2005spike, narisetty2014bayesian} and shrinkage priors \citep{park2008bayesian, carvalho2010horseshoe} under a Bayesian framework; we use the former. 
Specifically, the density of $\beta^j_{st}$ is given by 
\small
\vspace{-0.8cm}
\begin{align} \label{eq:spikeslab}
  \beta^j_{st} & \sim g^j_{st}N(0, \eta_1\tau_{s}^j) + (1-g^j_{st}) N(0, \eta_0\tau_{s}^j),
\end{align}
\vskip -0.8cm
\normalfont
\noindent where $g^j_{st}$ is defined in \S \ref{ggw} and $\eta_1\gg\eta_0>0$. The graphical prior is completed with a conjugate inverse-gamma prior, $\tau_s^j\sim \text{IG}(a_1,a_2)$ and the same Bernoulli prior on $g_{st}^j$ as in \S \ref{ggw}.


The pseudo-likelihood approach provides a fast and flexible way in estimating GGMs, especially when the Gaussian assumption cannot be verified. \citet{atchade2019quasi} showed that its posterior contraction rate matches the rate of convergence of the frequentist neighborhood selection. Alluding to the slightly faster upper bound on the rate of posterior contraction, \citet{atchade2019quasi} remarked that the pseudo-likelihood approach is statistically more efficient than a full likelihood approach (e.g., Gauusian-G-Wishart approach) for high-dimensional graphs.
Although we do not consider high-dimensional graphs ($q>n$) in this paper, the pseudo-likelihood approach is computational efficient as well. Its computation is considerably faster than the Gaussian-G-Wishart model because (i) there is no intractable normalizing constant in the graphical prior anymore, and (ii) computation of independent linear regression models are trivially parallelizable. To exploit parallel computing, we do not constrain $g_{st}^j=g_{ts}^j$. This asymmetry can be fixed with post-MCMC processing, which will be detailed in \S \ref{fix}.

\subsection{Partition averaging and graph prediction} \label{pagp}
Recall that by introducing the random partition $\bm{\rho}$, we effectively discretize the continuous precision matrix function $\Omega(\cdot)$ as  $\Omega(\mathrm{x}_i)=\Omega_j$ for $i\in S_j$, which greatly simplifies the modeling of covariate-dependent sparse precision matrix. Because $\bm{\rho}$ is random, we can recover $\Omega(\cdot)$ (hence $G(\cdot)$) by {\it partition averaging} the posterior distribution. 

\vspace{-0.4cm}
\subsubsection{Partition averaging} 
\vspace{-0.2cm}
The full posterior distribution is given by 
\small
\vspace{-0.75cm}
\begin{align*}
    p\{\Omega(\cdot), G(\cdot),\bm{\rho}\mid \mathrm{y}_{1:n},\mathrm{x}_{1:n}\}\propto \mathcal{P}\{\mathrm{y}_{1:n}\mid\Omega(\cdot)\}\cdot p\{\Omega(\cdot)\mid \bm{\rho},G(\cdot)\}\cdot p\{G(\cdot)\mid\bm{\rho}\}\cdot p(\bm{\rho}\mid\mathrm{x}_{1:n}),
\end{align*}
\vskip -0.75cm
\normalfont
\noindent where $\mathrm{y}_{1:n}$ and $\mathrm{x}_{1:n}$ denote $\mathrm{y}_1,\ldots,\mathrm{y}_n$ and $\mathrm{x}_1,\ldots,\mathrm{x}_n$, respectively. Marginalizing out partition $\bm{\rho}$,
\small
\vspace{-0.75cm}
\begin{align*}
    p\{\Omega(\cdot), G(\cdot)\mid \mathrm{y}_{1:n},\mathrm{x}_{1:n}\}&=\sum_{\bm{\rho}\in \mathcal{B}_n}p\{\Omega(\cdot), G(\cdot),\bm{\rho}\mid \mathrm{y}_{1:n},\mathrm{x}_{1:n}\}\\
    &\propto \sum_{\bm{\rho}\in \mathcal{B}_n}\mathcal{P}\{\mathrm{y}_{1:n}\mid \Omega(\cdot)\}\cdot p\{\Omega(\cdot)\mid \bm{\rho},G(\cdot)\}\cdot p\{G(\cdot)\mid \bm{\rho}\}\cdot p(\bm{\rho}\mid \mathrm{x}_{1:n})\\
    &=\sum_{\bm{\rho}\in \mathcal{B}_n}\Bigg[\mathcal{P}(\mathrm{y}_{1:n}\mid\Omega_1,\dots,\Omega_{K_{\bm{\rho}}})
    \prod_{j=1}^{K_{\bm{\rho}}}\Big\{p(\Omega_j\mid G_j)p(G_j)\Big\}\Bigg]
    p(\bm{\rho}\mid\mathrm{x}_{1:n}).
\end{align*}
\vskip -0.5cm
\normalfont
\noindent In summary, (i) through Bayesian hierarchical formulation via random partition, we reduce an infinite-dimensional quantity $\Omega(\cdot)$ as a sparse positive-definite matrix function of $\mathrm{x}$ to a finite number of sparse positive-definite matrices $\Omega_j$'s; and (ii) through partition averaging, we recover the infinite dimensional functional $\Omega(\cdot)$ from $\Omega_j$.

\subsubsection{Graph prediction}
The PxG model also allows for graph structure prediction given a new sample covariate  $\mathrm{x}_\text{new}$ through the posterior predictive distribution 
\small
\vspace{-0.8cm}
\begin{align}
    & p\{\Omega(\mathrm{x}_\text{new}),G(\mathrm{x}_\text{new})\mid \mathrm{y}_{1:n},\mathrm{x}_{1:n},\mathrm{x}_{\text{new}}\} \nonumber \\ 
    & =\sum_{\bm{\rho}\in\mathcal{B}_{n}}\sum_{j=1}^{K_{\max}}p(\mathrm{x}_{\text{new}}\in S_j\mid\bm{\rho}, \mathrm{y}_{1:n}, \mathrm{x}_{1:n})\cdot p(\Omega_j, G_j\mid\bm{\rho},\mathrm{y}_{1:n},\mathrm{x}_{1:n})\cdot p(\bm{\rho}\mid\mathrm{y}_{1:n},\mathrm{x}_{1:n} ), \label{eqn:gp}
\end{align} 
\vskip -0.5cm
\normalfont 
\noindent where $K_{\max}$ is the maximum number of clusters allowed.
This is a new and useful feature as the prediction does not require the access of $\mathrm{y}_{\text{new}}$.
Notice that the marginalization performed here cannot be done in non-Bayesian partition-based graphical models (e.g., \citet{liu2010graph}) since the partition is not random. 

\subsection{Theoretical properties of the PxG model}
In the following, we first investigate the flexibility of our prior specification for the PxG model, following which we shall investigate the asymptotic properties of the posterior distribution. 
\subsubsection{Prior support}\label{priorsupport}

Essential to studying the theoretical support of our prior is to look at an \textcolor{black}{equivalent} representation of the conditional density $p(\my \mid \mx)$, where $(\my, \mx)$ denotes a generic response-covariate pair. 
Based on the previous definition in \S 2, $\bm{\Omega}$ is the collection of  $q \times q$ precision-matrix atoms. Let $\bm{\mu} = \{\bm{\mu_j} \}_{j=1}^{K_{\bm\rho}}$ and $\bm{\Sigma^x} = \{\Sigma^x_j \}_{j=1}^{K_{\bm\rho}}$ denote the collection of normal mean-covariance atoms. For simplicity, in this subsection we assume $\Sigma^x_j = \sigma_j^2 I_p$. 
As in (\ref{eq:ylikelihood2}), an equivalent representation of the joint density of $\mathrm{y}$ and $\mathrm{x}$ is given by
\vskip -0.8cm
\small
\begin{equation*}
p(\my, \mx \mid \bm{\pi}, \bm{\Omega}, \bm{\mu}, \bm{\Sigma^x}) = \sum_{j=1}^{K_{\bm\rho}} \pi_j \mbox{N}_q(\my;  0, \Omega_j^{-1}) \mbox{N}_p(\mx;  \bm{\mu}_j, \sigma_j^2 I_p). 
\end{equation*}
\normalfont
\vskip -0.3cm
\noindent By integrating out $\mathrm{y}$, the marginal density of $\mx$ is
\vskip -0.8cm
\small
\begin{equation*}
p(\mx \mid \bm{\pi}, \bm{\mu}, \bm{\Sigma^x}) = \sum_{j=1}^{K_{\bm\rho}} \pi_j\mbox{N}_p(\mx; \bm{\mu}_{j}, \sigma_j^2 I_p),
\end{equation*}
\normalfont
\vskip -0.3cm
\noindent  Hence, we consider the specific form of the conditional density of the PxG model,
\vspace{-0.5cm}
\small
\begin{eqnarray}\label{eq:cond}
p(\my \mid \mx, \bm{\Omega}, \bm{\mu}, \bm{\Sigma^x}, \bm{\pi}) =  \frac{p(\my, \mx \mid \bm{\Omega}, \bm{\mu}, \bm{\Sigma^x}, \bm{\pi}) }{p(\mx \mid \bm{\mu}, \bm{\Sigma^x}, \bm{\pi})} = \sum_{j=1}^{K_{\bm\rho}} \pi_j(\mx) \mbox{N}_q(\my;  0, \Omega_j^{-1}),
\end{eqnarray}
\vskip -0.5cm
\normalfont
\noindent where 
\small
\vspace{-0.75cm}
\begin{eqnarray}\label{eq:wt}
\pi_j(\mx) = \frac{\pi_j\mbox{N}_p(\mx;  \bm{\mu}_j, \sigma_j^2 I_p)} {\sum_{t=1}^{K_{\bm\rho}} \pi_t\mbox{N}_p(\mx; \bm{\mu}_{t}, \sigma_t^2 I_p)}. 
\end{eqnarray}
\vskip -0.5cm
\normalfont
\noindent Equation \eqref{eq:cond} is a conditional density representation with covariate-dependent weights in \eqref{eq:wt}. Interestingly, if $\sigma_j^2 \equiv \sigma^2$, $\pi_j(\mx)$ depends only on linear functions of $\mx$, noting the simplification,
\small
\vspace{-0.5cm}
\begin{eqnarray}\label{eq:ml}
\pi_j(\mx) = \frac{\pi_j \exp\{(-\|\bm{\mu}_j\|^2/2+ \bm{\mu}_j^T\mx)/\sigma^2\}} {\sum_{t=1}^{K_{\bm\rho}} \pi_t \exp\{(-\|\bm{\mu}_t\|^2/2+ \bm{\mu}_t^T\mx)/\sigma^2\} },
\end{eqnarray}
\vskip -0.5cm
\normalfont
\noindent where $\|\cdot\|$ is the vector Euclidean norm. Equation \eqref{eq:ml} is reminiscent of multinomial-logit probabilities as one can reparameterize $\pi_j(\mx)$ in terms of the canonical parameters $(a_j, \bm{b}_j) \in \mathbb{R} \times \mathbb{R}^p$ of the Gaussian exponential family
\small
\vspace{-0.75cm}
\begin{eqnarray*}\label{eq:ml2}
\pi_j(\mx) = \frac{\pi_j \exp( a_j + \bm{b}_j^T\mx)} {\sum_{t=1}^{K_{\bm\rho}} \pi_t \exp(a_t +  \bm{b}_t^T\mx) }.
\end{eqnarray*}
\vskip -0.5cm
\normalfont
\noindent Thus, the PxG model with $\sigma_j^2 \equiv \sigma^2$ can be viewed as an extension of the HME models by \cite{jordan1994hierarchical}. \cite{norets2010approximation} obtained approximation results of an arbitrary continuous conditional density using the HME models of the form \eqref{eq:ml} under suitable regularity conditions, demonstrating the flexibility of such models. \cite{norets2017adaptive} took this a step further and obtained minimax rates of posterior convergence for appropriately smooth conditional densities. 

As $\my \mid \mx, \bm{\Omega}, \bm{\mu}, \bm{\Sigma^x}, \bm{\pi}$ is not Gaussian,  the variance-covariance matrix $\V(\my \mid \mx, \bm{\Omega}, \bm{\mu}, \bm{\Sigma^x}, \bm{\pi})$ no longer captures the conditional dependency relations among $(y_1, \ldots, y_q)$. \textcolor{blue}{Observe that the conditional independency relation here  holds true only by further conditioning on the latent clusters. Similar examples of conditional independency conditional on certain latent parameters can be found in the literature on robust graphical modeling \citep{finegold2011robust, finegold2014robust, cremaschi2019hierarchical}.} It nevertheless is important to look at the form of $\V(\my \mid \mx, \bm{\Omega}, \bm{\mu}, \bm{\Sigma^x}, \bm{\pi})$ and how it changes with covariates.  Observe that 
\small
\vspace{-0.7cm}
\begin{eqnarray*} 
\V(\my \mid \mx, \bm{\Omega}, \bm{\mu}, \bm{\Sigma^x}, \bm{\pi}) = \sum_{j=1}^{K_{\bm\rho}} \pi_j(\mx) \Omega_j^{-1}. 
\end{eqnarray*}
\vskip -0.7cm
\normalfont
Notice that the precision matrix  of $\my \mid \mx, \bm{\Omega}, \bm{\mu}, \bm{\Sigma^x}, \bm{\pi}$ is given by $\{\sum_{j=1}^{K_{\bm\rho}} \pi_j(\mx) \Omega_j^{-1}\}^{-1}$.  Given $\my$ and $\mx$, one can find a simplified representation of the posterior mean of $\V(\my \mid \mx, \bm{\Omega}, \bm{\mu}, \bm{\Sigma^x}, \bm{\pi})$ which is $\sum_{j=1}^{K_{\bm\rho}} \bbE \{\pi_j(\mx) \Omega_j^{-1} \mid \my, \mx\}$.
It is hard to characterize the nature and sparsity structure of $\{\sum_{j=1}^{K_{\bm\rho}} \pi_j(\mx) \Omega_j^{-1}\}^{-1}$ in its full generality, but if $\pi_j(\mx)$ behaves approximately like an indicator function $\ind_{\cX_j}(\mx)$ for a finite partition $\{\cX_1, \ldots, \cX_{K_{\bm\rho}}\}$ of $\cX\subseteq\mathbb{R}^p$, then the precision matrix of $\my \mid \mx$ would be approximately $\sum_{j=1}^{K_{\bm\rho}} \ind_{\cX_j}(\mx)\Omega_j$. 
In this case the precision matrix approximately reflects the conditional independence structure of $\my \mid \mx$. 
Theorem \ref{prop:suppPxG} below shows that if indeed all entries of the true conditional variance $\mbox{Var}(\my \mid \mx, \bm{\Omega}, \bm{\mu}, \bm{\Sigma^x}, \bm{\pi})$ is $\nu$-H\"{o}lder continuous, $0<\nu\leq 1$ (see the definition of $\nu$-H\"{o}lder continuity in the lemma below), then the prior assigns positive probability to appropriate integrated neighborhood  of the true function of the covariance matrix $\Omega_0^{-1}(\cdot)$. For simplicity, we only consider the case when the covariate $\mx$ is scalar for the following lemma and theory in this section and use notation $x$ for scalar $\mx$. Before introducing Theorem \ref{prop:suppPxG}, we first show the fact that $\nu$-H\"{o}lder continuous matrix can be approximated by a piecewise constant function.

\begin{lemma} \label{lemma}
  Denote the true conditional covariance matrix of $\mathrm{y}$ given $x$ as $\Omega_0^{-1}(x)=[\sigma^0_{ij}(x)]_{q\times q}$. Let $\mathcal{H}^\nu$ be the space of uniformly $\nu$-H\"{o}lder continuous functions, $0<\nu\leq 1$, i.e., 
  {\small
  \vspace{-0.5cm}
  \begin{equation*}
    \mathcal{H}^\nu = \Big\{f: [a,b]\rightarrow \mathbb{R};\,\, \|f\|_{\mathcal{H}^\nu} \equiv \sup_{x,x'\in[a,b]} \frac{|f(x)-f(x')|}{|x-x'|^\nu}<\infty \Big\}, 
  \end{equation*}
  \vskip -0.5cm
  \normalfont}
  \noindent where $\|f\|_{\mathcal{H}^\nu}$ is the $\nu$-H\"{o}lder coefficient.  Assume $\sigma^0_{ij}(x)\in\mathcal{H}^\nu$, $i,j=1,\ldots,q$. Therefore, for any $\epsilon>0$, there exists $K_\epsilon\in\mathbb{N}^+$ and a set of positive definite matrices $\{\Omega_{0k}\}_{k=1}^{K_\epsilon}$ where $\Omega_{0k}^{-1}=(\sigma^{0k}_{ij})_{q\times q}$ and a set of positive real numbers $a\leq a_1< \ldots < a_{K_\epsilon+1}\leq b$ such that
  {\small
  \vspace{-0.5cm}
  \begin{equation*}
    \Big\|\Omega_0^{-1}(\cdot) - \sum_{j=1}^{K_{\epsilon}}\ind_{(a_j,a_{j+1}]}(\cdot)\Omega_{0j}^{-1} \Big\|_1<\epsilon,
  \end{equation*}
  \vskip -0.5cm
  \normalfont}
 \noindent where $\Omega_{0j}^{-1}$ and $a_j$ depend only on $\Omega_0^{-1}(\cdot)$ and $\epsilon$, $\|A (\cdot)\|_1 = \int_a^b \|A(x)\|dx$ for any matrix norm $\| \cdot \|$.
\end{lemma}
\begin{proof}
    See Supplementary Materials \ref{sec:lemma}.
\end{proof}

\begin{theorem}\label{prop:suppPxG} {\normalfont(Prior large support).}
Assume that the true conditional density of $\my$ given $x$ satisfies $\mbox{Var}(\mathrm{y}\mid x)=\Omega_0^{-1}(x)$, where the precision matrix $\Omega_0(\cdot)$ is a continuous function of a one-dimensional covariate $x \in (a, b]$ for $-\infty<a<b<\infty$. Assume that all entries of the true covariance matrix $\Omega_0^{-1}(x)$ are in $\mathcal{H}^\nu$, $0<\nu\leq 1$.
Then there exists $K'_\epsilon\in\mathbb{N}^+$ such that for any absolutely continuous prior distributions on $(\Omega_j, \mu_j,\Sigma^x_j) \in \mathcal{S}_{q \times q}^+ \times \mathbb{R} \times \mathbb{R}^+$ and   $(\pi_1, \ldots, \pi_{K'_\epsilon})$ in the simplex $\Delta^{K'_\epsilon-1}$, 
{\small
\vspace{-0.2cm}
\begin{equation*}
  \Pi \Big \{ \big\| \Omega_0^{-1}(\cdot) - \sum_{j=1}^{K'_{\epsilon}}\pi_j(\cdot)\Omega_j^{-1} \big\|_1 < \epsilon \Big \} > 0,
\end{equation*}
\vskip -0.25cm
\normalfont
\noindent}
where $\mathcal{S}_{q \times q}^+$ is the cone of $q\times q$ positive definite matrices and $\pi_j(\cdot)$ is in the form of \eqref{eq:wt}.
\end{theorem}
This specific choice of $\nu$-H\"{o}lder continuous true conditional variance is  motivated by the good performance of the simulation study in \S \ref{sec:simu1} and \S \ref{sec:simu2}. 
Refer to Supplementary Materials \ref{sec:proof} for a proof of Theorem \ref{prop:suppPxG}.  Observe that the proof proceeds in two steps: i) Invoking Lemma \ref{lemma}, we first approximate a $\nu$-H\"{o}lder continuous using a piecewise constant function. ii) Approximating a piecewise constant density using densities of type \eqref{eq:cond}. This is in contrast with  \cite{norets2010approximation} and \cite{norets2017adaptive} who considered approximating only smooth conditional densities. Another difference with \cite{norets2010approximation} and \cite{norets2017adaptive} is that we obtain prior support in an integrated metric whereas \cite{norets2010approximation} obtain pointwise approximation bounds. 
Observe that for $\nu \in (0, 1]$ we approximate the $\nu$-H\"{o}lder covariance function with a piecewise constant function in an integrated metric. This does not hold for $\nu>1$ in a stronger metric (such as the supremum norm). Refer to Theorem 2 of \citet{castillo2014bayesian}, which requires $\alpha > 1$ for the supremum-norm posterior contraction results to hold.



One of our key contributions is to recognize the connection between conditional density in (\ref{eq:cond}) to the mixture of experts representation in \cite{norets2017adaptive} under a special choice of parameters. This helps us adapt results on prior large support, and later in \S 3.4.2 allows us to leverage on the posterior contraction rate for the joint density \citep{shen2013adaptive}.

In addition to the large prior support property, we also develop posterior contraction rates in \S \ref{ssec:postcon} where it is important to assume smoothness.  In fact, as we shall see below in \S \ref{ssec:postcon}, our derivation for rate of contraction assumes true joint density to be a  mixture of Gaussians. As shown in  \cite{ghosal2000convergence}, this is a fairly broad class of densities and can approximate any smooth density with desirable accuracy.  In addition, we characterize  the role of the number of mixture components $K$ in rate, but assume the dimension to be fixed. 


\subsubsection{Posterior contraction rates}\label{ssec:postcon}
In this section, we investigate the asymptotic properties of the posterior distribution associated with the PxG model.  For technical simplicity, we shall consider a fractional likelihood introduced in \citet{bhattacharya2019bayesian}, although we anticipate analogous results with the original likelihood. Let $p_{\Theta}$ be the joint density of $\mathrm{x}$ and $\mathrm{y}$ in the PxG model and $p_{\Theta_0}$ be the true model, where $\Theta$ and $\Theta_0$ are the model parameters and the true parameters, respectively. Instead of fitting $p_\Theta$, we shall work with 
$p_\Theta^\alpha$ for $\alpha \in (0, 1)$ and investigate the convergence rate of the posterior $\Pi_{n, \alpha}$ given by 
\small
\vspace{-1.6cm}
\begin{eqnarray*}
\Pi_{n, \alpha}(p_\Theta \,\,\big|\,\, \mathrm{x}_{1:n},\mathrm{y}_{1:n}) = 
 \frac{p_\Theta^\alpha(\mathrm{x}_{1:n},\mathrm{y}_{1:n}) \Pi(\Theta)}{\int p_\Theta^\alpha(\mathrm{x}_{1:n},\mathrm{y}_{1:n}) \Pi(\Theta) d\Theta}
\end{eqnarray*}
\vskip -0.7cm
\normalfont
\noindent
for some prior $\Pi$ on the parameters $\Theta$.
Next we shall state two types of assumptions on the true density $p_{\Theta_0}$ and assumed prior $\Pi$, respectively. 


\begin{assumption}\label{assump:true} {\normalfont(Assumptions on the conditional density and marginal density of covariates).} Assume the true conditional density of the $q$-dimensional response $\mathrm{y}$ given the $p$-dimensional covariate $\mathrm{x}$ has the following form
{\small
\vspace{-0.5cm}
\begin{align*}
    p(\mathrm{y} \mid \mathrm{x}, \Theta_0)
     = \sum_{j=1}^K \pi_{j}(\mathrm{x}\mid \Theta_0) \cdot \mbox{N}_q(\mathrm{y};\bm{0},\Sigma^0_{G_{0j}}) 
     = \sum_{j=1}^K \pi_{j}(\mathrm{x}\mid \Theta_0) \cdot \phi_{\Sigma^0_{G_{0j}}},
\end{align*}
\vskip -0.5cm
\normalfont
\noindent}
and the true marginal density of covariates is
{\small
\vspace{-0.5cm}
\begin{align*}
    p(\mathrm{x}\mid \Theta_0) = \sum_{j=1}^K \pi_{0j}\cdot \mbox{N}_p (\mathrm{x};\bm{\mu}_{0j}, \Sigma^x_{0j})
    =  \sum_{j=1}^K \pi_{0j}\cdot \phi_{\bm{\mu}_{0j}, \Sigma^x_{0j}},
\end{align*}
\vskip -0.5cm
\normalfont
\noindent}
where 
{\small
\vspace{-0.5cm}
\begin{equation*}
    \pi_{j}(\mathrm{x}\mid \Theta_0) = \frac{\pi_{0j}\cdot \phi_{\bm{\mu}_{0j}, \Sigma^x_{0j}} }{\sum_{j=1}^K \pi_{0j}\cdot \phi_{\bm{\mu}_{0j}, \Sigma^x_{0j}}},
\end{equation*}
\normalfont}
\noindent
$\Theta_0=\{ (\pi_{0j}, \Sigma^0_{G_{0j}},\bm{\mu}_{0j},\Sigma^x_{0j}) \}_{j=1}^K$,
$K$ is a known constant and $\Sigma^x_{0j}=\mbox{diag}(\sigma^2_{0j1},\ldots,\sigma^2_{0jp})$ is a diagonal matrix with diagonal entries $\sigma^2_{0ji}$, $i=1,\ldots, p$, $j=1,\ldots,K$. Let $p_{\Theta_0} = p(\mathrm{y}\mid \mathrm{x},\Theta_0)\cdot p(\mathrm{x}\mid\Theta_0)$ be the true joint density and $p_\Theta$ be any joint density with parameters $\Theta=\{ (\pi_{j}, \Sigma_{G_{j}},\bm{\mu}_{j},\Sigma^x_{j}) \}_{j=1}^K$, where $\bm{\mu}_j=(\mu_{j1},\ldots,\mu_{jp})$ and $\Sigma^x_j=\mbox{diag}(\sigma^2_{j1},\ldots,\sigma^2_{jp})$. Moreover, let $0<\lambda^0_{j1}\leq\ldots\leq\lambda^0_{jq}$ be the eigenvalues of $\Sigma^0_{G_{0j}}$. Assume $0<\lambda^0_m\leq\lambda^0_{j1}\leq\ldots\leq\lambda^0_{jq}\leq\lambda^0_M$ and  $\sigma_m^2\leq\sigma^2_{0ji}\leq\sigma_M^2$, for $j=1,\ldots,q$, $i=1,\ldots,p$, where $\lambda^0_m$, $\lambda^0_M$, $\sigma_m^2$, $\sigma_M^2$ are global constants.
\end{assumption}

\begin{assumption}\label{assump:prior} {\normalfont (Assumptions on $\Pi$).} \\
    (1) Prior for $(\pi_1,\ldots,\pi_K)$ is Dirichlet $(\pi_0/K,\ldots,\pi_0/K)$, $\pi_0>0$. \\
    (2) Prior for $\mu_{ji}$ is any normal density, $j=1,\ldots,K$, $i=1,\ldots,p$. \\
    (3) Prior for $\sigma^2_{ji}$ is any inverse-Gamma density, $j=1,\ldots,K$, $i=1,\ldots,p$. \\
    (4) Prior for $\Sigma_{G_j}$ given $G_j$ is any G-Wishart density and edges in $G_j$ follow independent Bernoulli density, $j=1,\ldots,K$.
\end{assumption}

To measure the closeness between $p_{\Theta_0}$ and $p_{\Theta}$, define the $\alpha$-divergence with respect to the true probability measure $\mathbb{P}_{\Theta_0}$ as
\small
\vspace{-0.75cm}
\begin{equation*}
    D_{p_{\Theta_0},\alpha}(p_{\Theta},p_{\Theta_0}):=\frac{1}{\alpha-1}\log\int \Big(\frac{p_{\Theta}}{p_{\Theta_0}}\Big)^\alpha p_{\Theta_0} d\mu.
\end{equation*}
\vskip -0.5cm
\normalfont
\noindent
Using the two assumptions above, we derive the main theorem of the fractional posterior distribution around the true density.

\begin{theorem}\label{th:contraction} {\normalfont(Contraction of fractional posterior distributions).} 
Fix $\alpha\in(0,1)$ and let $\epsilon_n = \sqrt{\frac{K(p+q)\log n}{n}}$. Suppose Assumptions \ref{assump:true} and \ref{assump:prior} hold. Then, for any $D\geq 2$ and $t>0$,
\small
\vspace{-0.5cm}
\begin{equation*}
    \Pi_{n,\alpha}\Big( D_{p_{\Theta_0},\alpha}(p_{\Theta},p_{\Theta_0})\geq \frac{D+3t}{1-\alpha}\epsilon_n^2 \,\,\Big|\,\, \mathrm{x}_{1:n},\mathrm{y}_{1:n} \Big) \leq e^{-t\epsilon_n^2}
\end{equation*}
\vskip -0.5cm
\normalfont
\noindent
holds with $\mathbb{P}_{\Theta_0}$ probability at least $1-2/\{(D-1+t)^2\epsilon_n^2\}$, where $ \Pi_{n,\alpha}(\cdot\mid \mathrm{x}_{1:n},\mathrm{y}_{1:n})$ is the posterior probability and $\mathrm{x}_{1:n},\mathrm{y}_{1:n}$ are the sample covariates and responses, respectively.
\end{theorem}
\begin{proof}
    See Supplementary Materials \ref{proof2}.
\end{proof}
Compared to the regular posterior contraction results, the fractional posterior result in Theorem \ref{th:contraction} only requires the  prior to be sufficiently concentrated around $\Theta_0$. Observe that we have worked with the simplifying assumption when the true density $p_{\Theta_0}$ is itself a mixture of $K$-Gaussians and our PxG model assumed the knowledge of the number $K$. Indeed the convergence rate $\sqrt{K(p+q)\log n / n}$ is minimax optimal \citep{tsybakov2008introduction}.  When the number of mixtures $K = n^{\frac{p+q}{2\beta+p+q}},\beta>0$, one can recover the contraction rates of posterior for estimating $(p+q)$-variate $\beta$-smooth densities \citep{norets2017adaptive, shen2013adaptive}.

\section{Simulation Study} \label{sec:simu}
In this section, we present three simulated examples. The first two simulation studies consider the entries of precision matrix to be linear or piecewise linear functions of covariates whereas the third simulation assume discrete functions (i.e., piecewise constant precision matrices). We demonstrate that the proposed PxG model is able to handle both types of precision matrix functions. To show the necessity of using covariate information, we compare the PxG model with two following ``partial'' models: covariate-only model where graph is assumed to be constant among sample; graph-only model where graph varies with samples but does not depend on covariates. Model comparison among them is conducted via DIC in this section.

\subsection{Example 1: piecewise linear precision matrix} \label{sec:simu1}
\begin{figure}[H] 
\centering
 \scalebox{1}{\includegraphics[width=15cm]{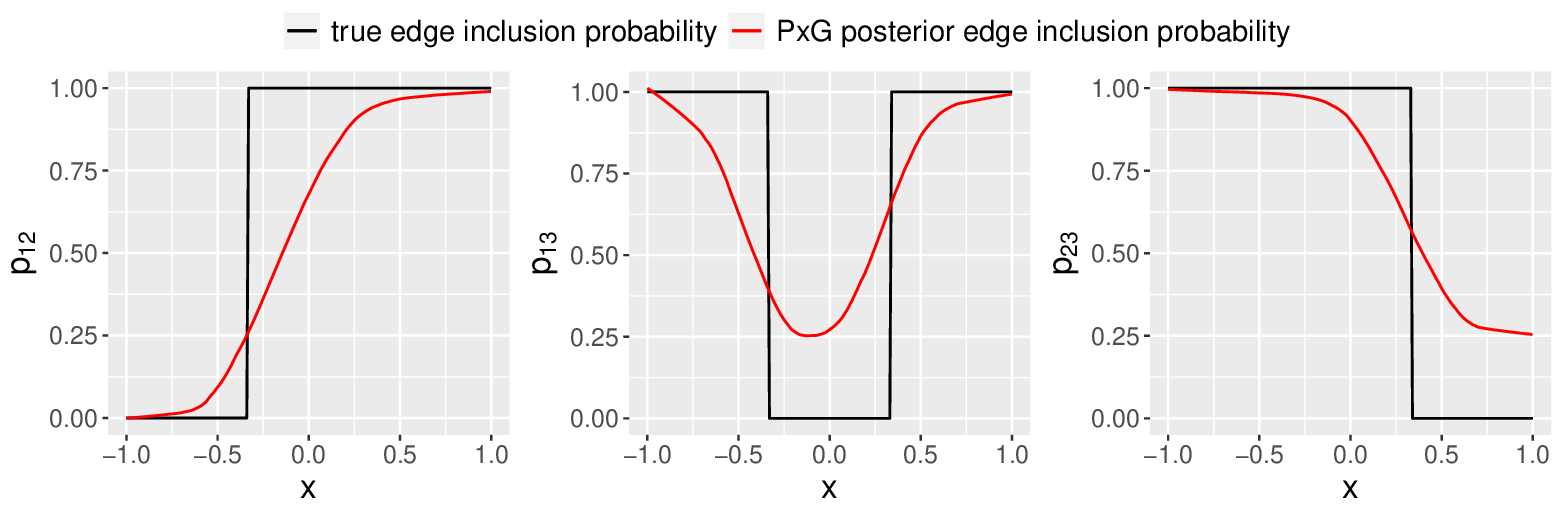}}
\caption{\small \relscale{1} Edge inclusion probability as a function of covariate $x$. Black: true edge inclusion probability. Red: PxG posterior estimate of edge inclusion probability.}\label{fig:sim1_1}
\end{figure}

\begin{figure}[H] 
\centering
\scalebox{1}{\includegraphics[width=15cm]{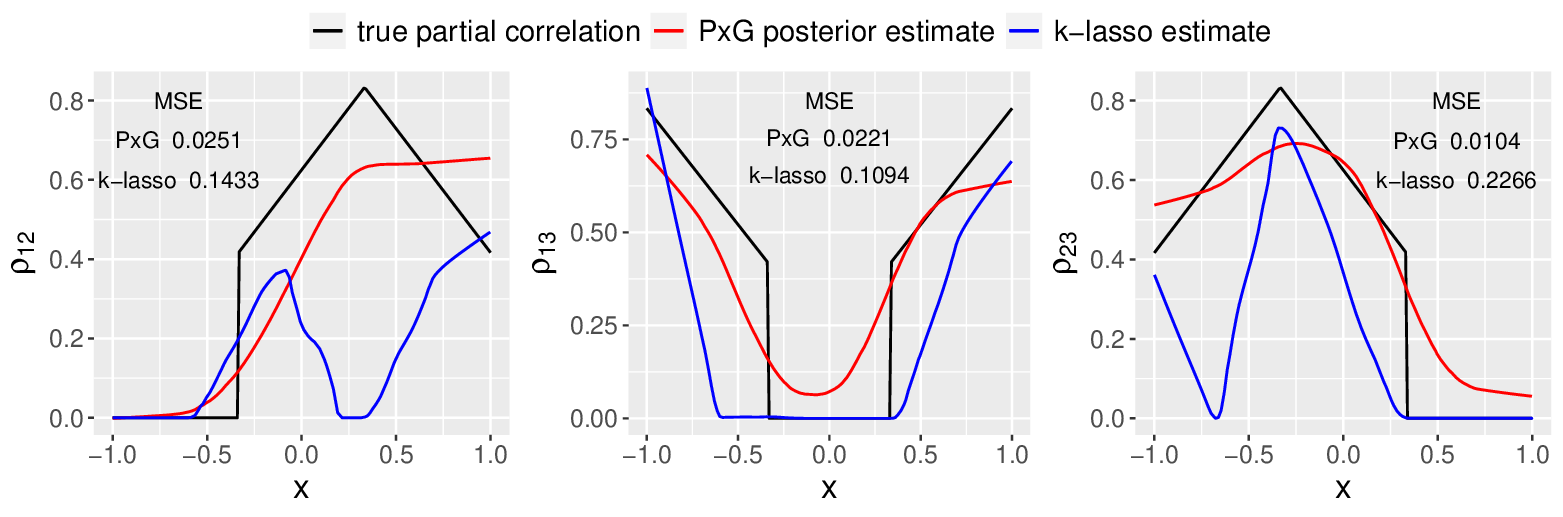}}
\caption{\small \relscale{1} Partial correlation as a function of covariate $x$.  Black: true partial correlation. Red: PxG  posterior estimate of partial correlation. Blue: k-lasso estimate of partial correlation.} \label{fig:sim1_2}
\end{figure}

We consider an example with a 3-dimensional GGM of which the precision matrix is a piecewise linear function of the given 1-dimensional covariate $x$. Define the true covariate-dependent precision matrix as $\Omega(x) = [\Omega_{st}(x)]_{s,t=1,2,3}$ where $\Omega_{ss}(x)=1.2$ and
\small
\vspace{-0.75cm}
\begin{align*}
    & -1<x<-0.33, \,\,\, \Omega_{12}(x) = 0, \,\,\, \Omega_{13}(x) = -0.75x+0.25, \,\,\, \Omega_{23}(x) = 0.75x+1.25; \\
    & -0.33\leq x<0.33, \,\,\, \Omega_{12}(x) = 0.75x+0.75, \,\,\, \Omega_{13}(x) = 0, \,\,\, \Omega_{23}(x) = -0.75x+0.75; \\
    & 0.33\leq x<1, \,\,\, \Omega_{12}(x) = -0.75x+1.25, \,\,\, \Omega_{13}(x) = 0.75x+0.25, \,\,\, \Omega_{23}(x) = 0.
\end{align*}
\vskip -0.75cm
\normalfont
\noindent Thus, the piecewise linear function of the precision matrix defines three clusters with different graphs. They are $G_1=(V,E_1)$ for $-1<x<-0.33$ with $E_1=\{1-3, 2-3\}$; $G_2=(V,E_2)$ $-0.33<x<0.33$ with $E_2=\{1-2, 2-3\}$; $G_3=(V,E_3)$ for $0.33\leq x<1$ with $E_3=\{1-2, 1-3\}$, where $V=\{1,2,3\}$. We draw 100 realizations of the covariate uniformly from each of the three clusters, i.e., the total sample size is 300. 
Then the responses $\mathrm{y}$ are drawn from the corresponding covariate-dependent normal distribution $N_3\{0,\Omega^{-1}(x)\}$. We simulate 1,000 MCMC samples using the blocked Gibbs sampler for the Gaussian-G-Wishart model with 500 burn-in samples. We plot the posterior edge inclusion probability as a function of $x$ for all three edges in Figure \ref{fig:sim1_1}. The black lines are the true edge inclusion and the red lines are the estimated posterior edge inclusion probabilities. The posterior edge inclusion probability is captured by averaging out all MCMC samples. We can also calculate the posterior partial correlation matrix as a function of the covariate $x$ based on the MCMC samples; see Figure \ref{fig:sim1_2}. From both figures, the posterior estimates capture the overall trends of the true edge inclusion and the true piecewise linear precision matrix. 

Moreover, we compare the PxG model with the kernel graphical lasso approach (hereafter, k-lasso) described in \citet{liu2010graph}, which estimates a covariate-dependent covariance matrix via kernel smoothing. Let $S(x)$ be the covariance matrix varying with a covariate $x$, then
\small
\vspace{-0.5cm}
\begin{equation*}
  S(x) = \sum_{i=1}^n K\Bigg(\frac{|x-x_i|}{h}\Bigg)(\mathrm{y}_i-\bm{\mu}(x))(\mathrm{y}_i-\bm{\mu}(x))^T\Big/ \sum_{i=1}^n K\Bigg(\frac{|x-x_i|}{h}\Bigg)
\end{equation*}
\normalfont
\noindent
with
\small
\begin{equation*}
  \bm{\mu}(x) = \sum_{i=1}^n K\Bigg(\frac{|x-x_i|}{h}\Bigg)\mathrm{y}_i\Big/ \sum_{i=1}^n K\Bigg(\frac{|x-x_i|}{h}\Bigg).
\end{equation*}
\vskip -0.1cm
\normalfont
\noindent
Here $h>0$ is the bandwidth and $K(\cdot)$ is a Gaussian kernel. We tune $h$ on a [0.1,1] grid with the Akaike information criterion.
To obtain a sparse precision matrix $\Omega_i$ for each sample, glasso \citep{friedman2008sparse} is applied,
$\hat{\Omega}_i = \arg \min_{\Omega} \big\{ \log|\Omega| - \mbox{tr}(S(x_i)\Omega)-\lambda_i\|\Omega\|_1 \big\}$.
The corresponding blue curves in Figure \ref{fig:sim1_2} shows the partial correlation estimates using k-lasso approach. We can see that it does not capture the basic shape of the piecewise linear partial correlation functions. And k-glasso approach does not yield posterior estimates of edge inclusion probabilities. Mean squared errors (MSE) of PxG and k-lasso for each partial correlation are shown in Figure \ref{fig:sim1_2} as well.
By comparing the proposed PxG model (DIC=2,193) with the covariate-only model (DIC=2,585) and the graph-only model (DIC=3,190) and via the deviance information criterion (DIC) (see Supplementary Materials for details), the PxG model has the smallest DIC value, indicating the best model fitting.

\subsection{Example 2: linear precision matrix with constant graph} \label{sec:simu2}
In this scenario, we demonstrate that the PxG model can recover the precision matrix as a function of covariates even when the graph induced by the precision matrix is constant. We compare PxG with approaches proposed by \citet{bhadra2013joint} (hereafter, BM13) and \citet{deshpande2019simultaneous} (hereafter, mSSL) to show the necessity of using covariate information in graph estimates and to show that under this case model misspecification leads to incorrect graph estimates. We consider a 5-dimensional covariate-dependent precision matrix $\Omega(x)$ as a linear function of a scalar covariate $x$. It is defined as follows: $\Omega_{ss}(x)=1.4$ for $s=1,2,3,4,5$; $\Omega_{s,s+1}(x)=x$ for $s=1,2,3,4$; the rest entries in $\Omega(x)$ are 0 and $x\in (-0.8,0)\cup(0,0.8)$. Thus, 4 entries in $\Omega(x)$ are linear functions of $x$. The constant graph induced by $\Omega(x)$ is a 5-node chain graph $G=(V,E)$ with 4 edges, where $V=\{1,2,3,4,5\}$ and $E=\{1-2,2-3,3-4,4-5\}$. Covariate $x$ is drawn uniformly from its range with sample size $n=300$, and the corresponding responses $\mathrm{y}$ are generated from $N_5\{0,\Omega^{-1}(x)\}$. We simulate 2,000 MCMC samples using the blocked Gibbs sampler of the Gaussian-G-Wishart model with 1,000 burn-in samples.

\begin{figure}[H] 
\centering
 \scalebox{1}{\includegraphics[width=16cm]{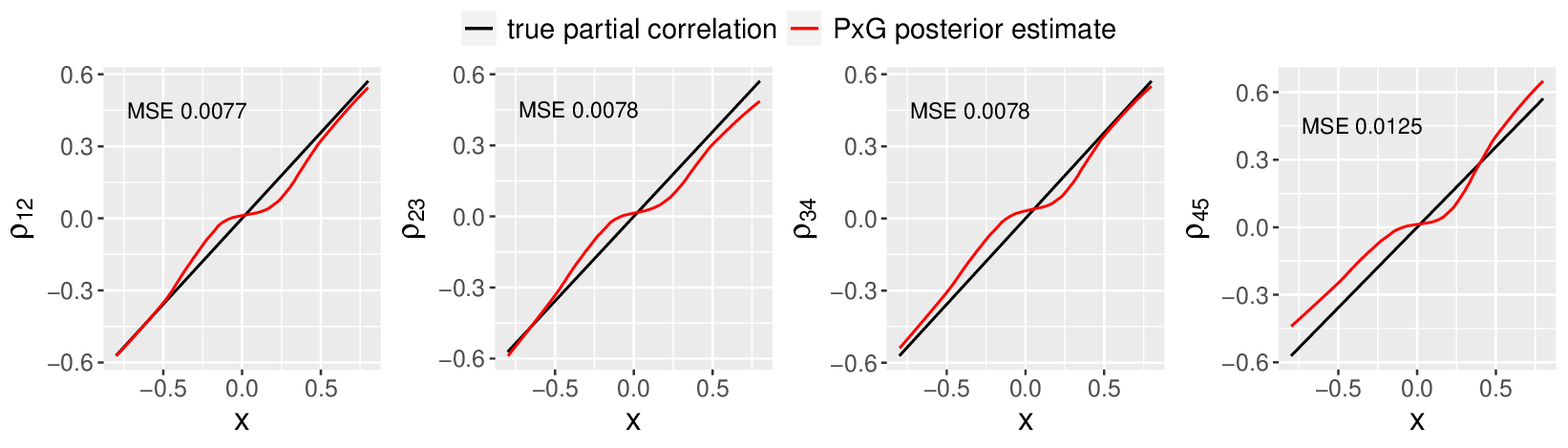}}
 \caption{\small \relscale{1} PxG posterior estimate of partial correlations.}\label{fig:sim2_1}
\end{figure}

\begin{figure}[H] 
\centering
 \scalebox{1}{\includegraphics[width=13cm]{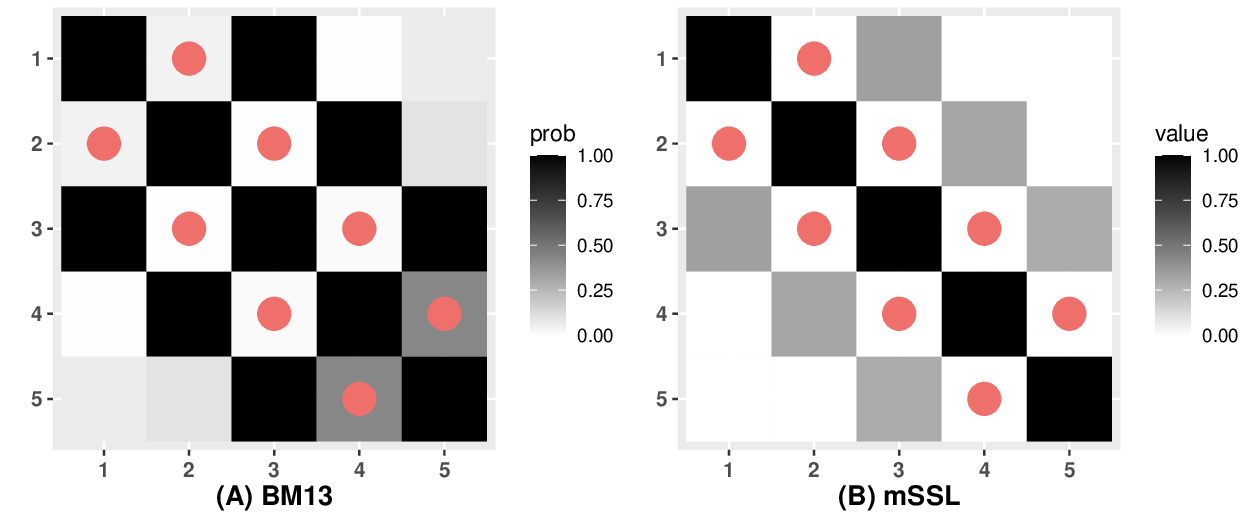}}
 \caption{\small \relscale{1} (A) BM13 posterior estimate of edge inclusion probability. (B) mSSL estimate of absolute value of partial correlation. Red dots indicate true edges in $\Omega(x)$.}\label{fig:sim2_2}
\end{figure}

We plot the posterior estimate of partial correlation as a function of $x$ for the four edges in Figure \ref{fig:sim2_1} indicated by red line along with the MSE for each partial correlation estimate. The black line is the true partial correlation. Thanks to partition averaging, the PxG model captures the linear precision matrix well even though all observations share the same graph and there is no cluster induced from the underlying data generating process. Figure \ref{fig:sim2_2} shows the results from BM13 and mSSL. For BM13, the posterior estimate of edge inclusion probability is shown in Figure \ref{fig:sim2_2} (A); for mSSL, we plot the absolute value of the estimate of partial correlation matrix in Figure \ref{fig:sim2_2} (B). Red dots in both plots are true edges. Both methods fail to identify the true edges induced by $\Omega(x)$ due to model misspecification since BM13 and mSSL assume a constant graph and a constant covariance matrix, both of which are independent from covariates. Without utilizing covariate information, the graph estimates from BM13 and mSSL seem to be dictated by the sample partial correlations of responses. In fact, we observe that sample partial correlations of edges $1-3$, $2-4$ and $3-5$ are about 4 times larger than the rest (including the true edges). 

\subsection{Example 3: piecewise constant precision matrix} \label{sec:simu3}
In this example, we consider the precision matrix to be a piecewise constant function of a set of covariates. Thus, clusters can be induced from the precision matrix. Assume $\Omega(\mathrm{x})$ to be a 50-dimensional covariate-dependent precision matrix where $\mathrm{x}$ is a 10-dimensional covariate. The piecewise constant precision matrix of $\Omega(\mathrm{x})$ is defined as $\Omega(\mathrm{x})=\Omega_1$ for $\mathrm{x}\sim N_{10}(\bm{0}_{10},I_{10})$ and $\Omega(\mathrm{x})=\Omega_2$ for $\mathrm{x}\sim N_{10}(\bm{2}_{10},I_{10})$, where $\bm{0}_{10}=0\cdot\bm{1}_{10}$ and $\bm{2}_{10} = 2\cdot\bm{1}_{10}$. Two clusters are induced by $\Omega(\mathrm{x})$, and $\Omega_1$ and $\Omega_2$ are the cluster-specific constant precision matrices. Let $G_1$ and $G_2$ be the corresponding sparse graphs induced from $\Omega_1$ and $\Omega_2$, respectively. They are selected randomly with  approximately 1\% sparsity, and then $\Omega_1$ and $\Omega_2$ are drawn from  G-Wishart distribution with the degree of freedom equal to 3 and scale matrix equal to the identity matrix given $G_1$ and $G_2$. We generate 250 realizations of covariate $\mathrm{x}$ for each cluster and the corresponding responses are drawn from $N_{50}\{\bm{0}_{10},\Omega^{-1}(\mathrm{x})\}$.

\begin{figure}[h] 
\centering
 \scalebox{1}{\includegraphics[width=16cm]{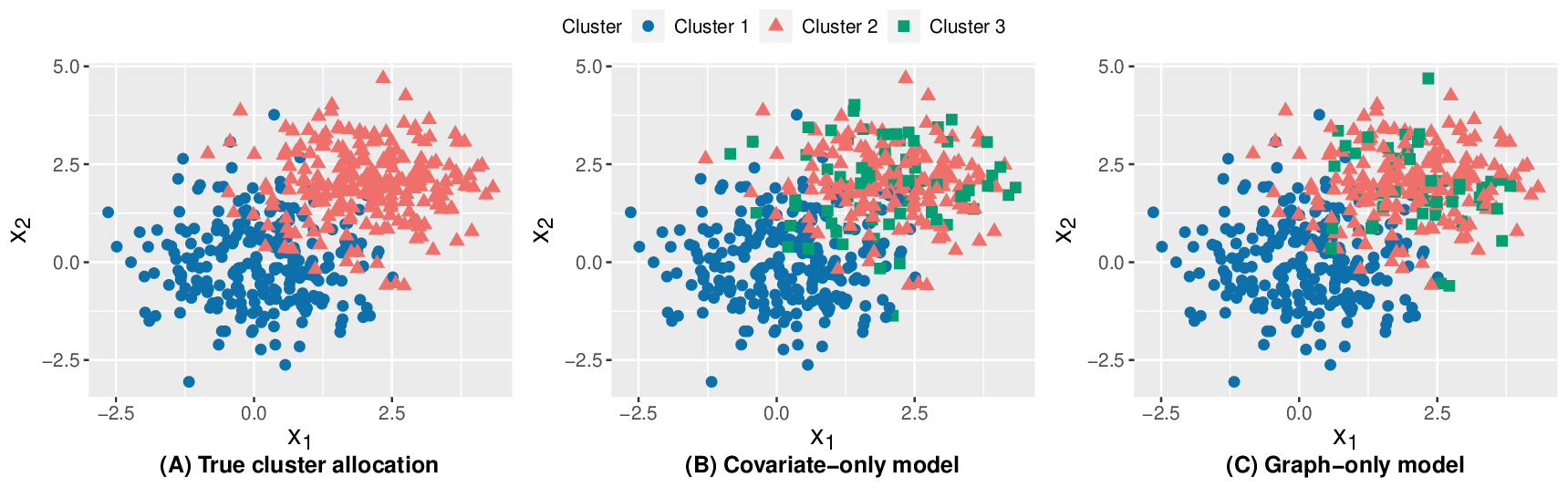}}
\caption{\small \relscale{1} (A) indicates the true cluster allocation, which is also the estimated cluster allocation of PxG (i.e., no misclassification). (B) and (C) are the posterior estimates of cluster allocations of the covariate-only model and the graph-only model, respectively.} \label{fig:sim3_1}
\end{figure}

Figure \ref{fig:sim3_1} (A) shows the first 2 dimensions of the 10-dimensional covariates with cluster allocation. The variances of both clusters are relatively large comparing to the distance between their means. Thus, we observe that there is significant overlap between these two clusters. Due to the relatively large dimension of graph in this example, we use the pseudo-likelihood model with the blocked Gibbs sampler. The Gaussian-G-Wishart model is too slow for such application due to the computational burden of calculating the normalizing constant of G-Wishart density. In the pseudo-likelihood approach, $g_{st}^j$ in (\ref{eq:spikeslab}) is not necessary the same as $g_{ts}^j$ since they are two separate parameters in two independent conditional likelihood functions that make up the pseudo-likelihood function. Therefore, the posterior edge inclusion probabilities of $g_{ts}^j$ and $g_{st}^j$, denoted as $p_{ts}^j$ and $p_{st}^j$, are unlikely to be the same. We use $\max(p_{ts}^j,p_{st}^j)$ as the final posterior edge inclusion probability for edge $s-t$ (similar in spirit to the union approach in the frequentist neighborhood selection method). This ensures the graph estimates are symmetric. We simulate 1,000 MCMC samples with 500 burn-in samples. For comparison, we follow the same procedure for the covariate-only model and the graph-only model.

\begin{figure}[h] 
\centering
 \scalebox{0.75}{\includegraphics[width=15cm]{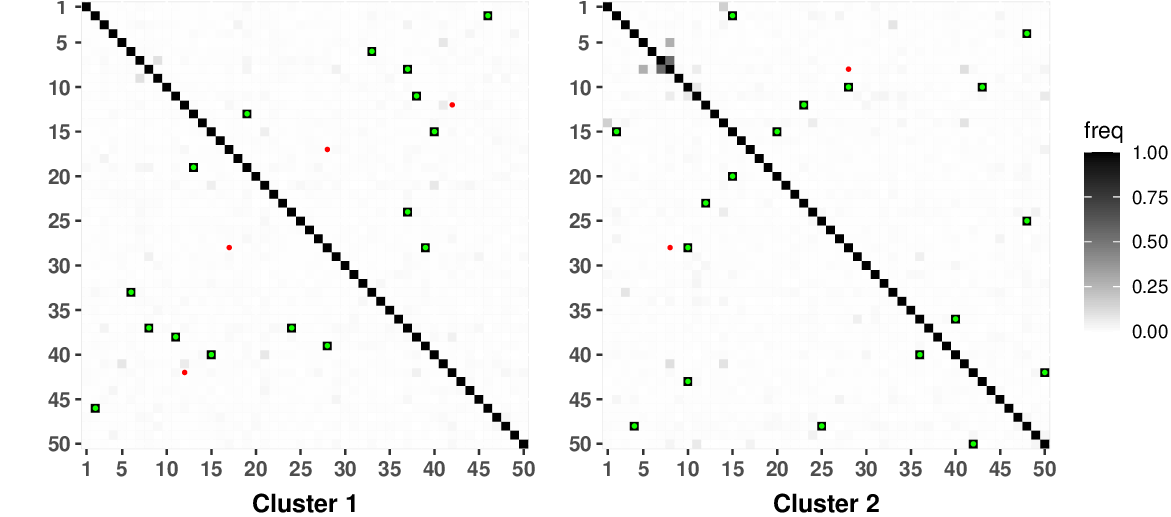}}
 \caption{\small \relscale{1} PxG posterior edge inclusion probability for Cluster 1 and 2. Red and green dots in both plots are the true edges in $G_1$ and $G_2$ with red indicating false negative edges and green for true positive edges.} \label{fig:sim3_2}
\end{figure}

\begin{figure}[h] 
\centering
 \scalebox{1.12}{\includegraphics[width=15cm]{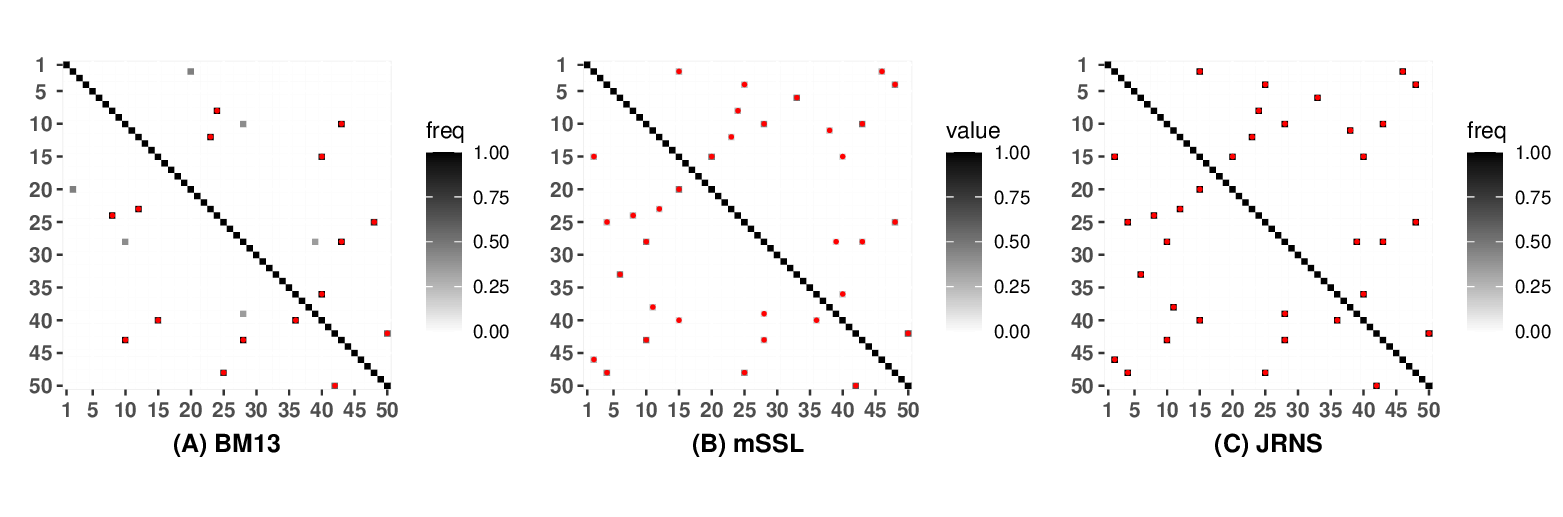}}
 \caption{\small \relscale{1} (A) BM13 posterior estimate of edge inclusion probability. (B) mSSL estimate of absolute value of partial correlation. (C) JRNS posterior estimate of edge inclusion probability. Red dots indicate edges selected by each approach.} \label{fig:sim3_3}
\end{figure}

The PxG model correctly identifies both clusters with no misclassification. Edge inclusion probabilities for each cluster are show in Figure \ref{fig:sim3_2}. Red and green dots are false negative edges and true positive edges respectively according to the true graph in each cluster. By selecting edges with posterior edge inclusion probabilities greater than 0.5, its posterior graph estimate of $G_1$ has 2 false negative edges and no false positive edges; for $G_2$ it has 1 false negative edge and 1 false positive edge. Due to the closeness of covariates between the two clusters, the covariate-only model is not able to correctly identify both clusters as shown in Figure \ref{fig:sim3_1} (B). Figure \ref{fig:sim3_1} (C) indicates that, without utilizing any covariate information, the response information is not enough for the graph-only model to identify Cluster 2 correctly, hence the true piecewise precision matrix as well. This demonstrates the necessity of using covariate information to assist graph estimates in heterogeneous GGMs. The DIC values for the PxG, covariate-only, and graph-only models are 104166, 155453, and 106138, respectively with the PxG having the smallest value among them.
We compare the PxG model with BM13, mSSL, and a recent approach by \citet{samanta2022generalized} (hereafter, JRNS) in this example. We plot the posterior edge inclusion probability for BM13 in Figure \ref{fig:sim3_3} (A), the estimates of absolute values of partial correlations using mSSL in Figure \ref{fig:sim3_3} (B), and the posterior edge inclusion probability for JRNS in Figure \ref{fig:sim3_3} (C) with red dots indicating edges selected by each approach. For BM13 and JRNS, the red dots are edges with posterior edge inclusion probability greater than 0.5 and for mSSL they are edges corresponding to nonzero entries in the estimate of precision matrix. Since the graph and the precision matrix are assumed to be constant among all observations, these three approaches suffer from model misspecification and do not identify the true edges in the graph. The graph estimates of mSSL and JRNS are identical, both having 17 edges. Compared with the true $G_2$, they have 1 false negative edge and 8 false positive edges. It is worse when compared to true $G_1$. The posterior graph estimate of BM13 has 8 edges whose posterior edge inclusion probabilities are greater than 0.5. It has 5 false negatives and 3 false positives compared with the true $G_2$. 




\section{Real Data Analysis: Breast Cancer Data} \label{sec:example}
In this section, we apply the proposed PxG model to a breast cancer dataset. The mRNA and protein data of breast cancer are downloaded from The Cancer Genome Atlas (TCGA) and the Clinical Proteomic Tumor Analysis Consortium (CPTAC) through the R program \texttt{TCGA-Assembler} \citep{wei2018tcga}. The central dogma of genetics states that genetic information flows from mRNA to protein. It is natural that we use mRNA expressions as covariates and protein expressions as responses in our approach. We model the protein network of breast cancer as a mRNA-dependent graph. The PxG model allows us to uncover how interactions between proteins change with different mRNA expression. Different cancer treatments take advantages of various properties of protein-protein interactions. Therefore, discovering how those interactions vary with mRNA expression is one of the key steps in developing new targeted, personalized cancer therapy.

It has been known that human epidermal growth factor receptor 2 (HER2), estrogen receptor (ER), and progesterone receptor (PR) are important in determining breast cancer subtypes \citep{onitilo2009breast}. They are related to treatment, survival rate, and cancer cell metabolism of breast cancer \citep{lobbezoo2013prognosis, rimawi2015targeting, kennecke2010metastatic}. We focus on the corresponding mRNAs of these three receptors, which are ERBB2 for HER2, ESR1 for ER and PGR for PR. For proteins, in addition to the protein expression of the three markers, we also include 12 proteins from the mTOR pathway, which is known to play a significant role in breast cancer \citep{hynes2006mtor, cidado2012targeting, mcauliffe2010deciphering}. In total, 883 observations are available for this analysis.

\begin{figure}[h] 
\centering
\scalebox{1}{\includegraphics[width=17cm]{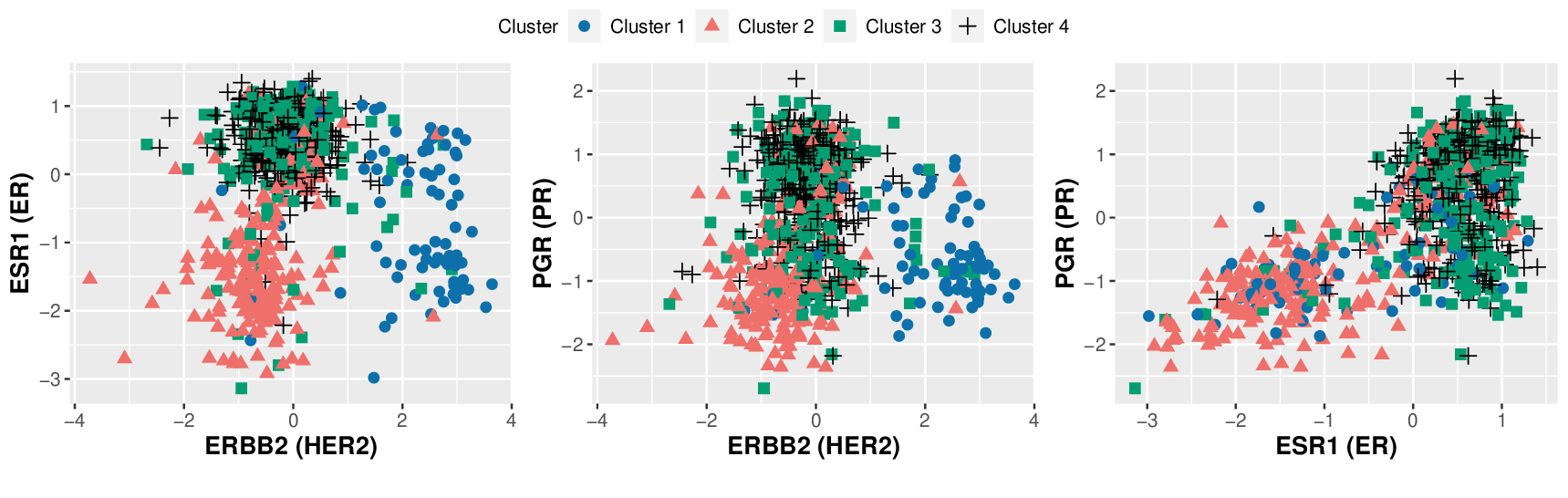}}
\caption{\small \relscale{1} Posterior estimates of cluster allocation of the PxG model shown on pairwise scatter plots of 3 mRNA covariates: ERBB2, ESR1 and PGR.} \label{fig:real_1}
\end{figure}

\begin{figure}[h] 
\centering
\scalebox{1}{\includegraphics[width=17cm]{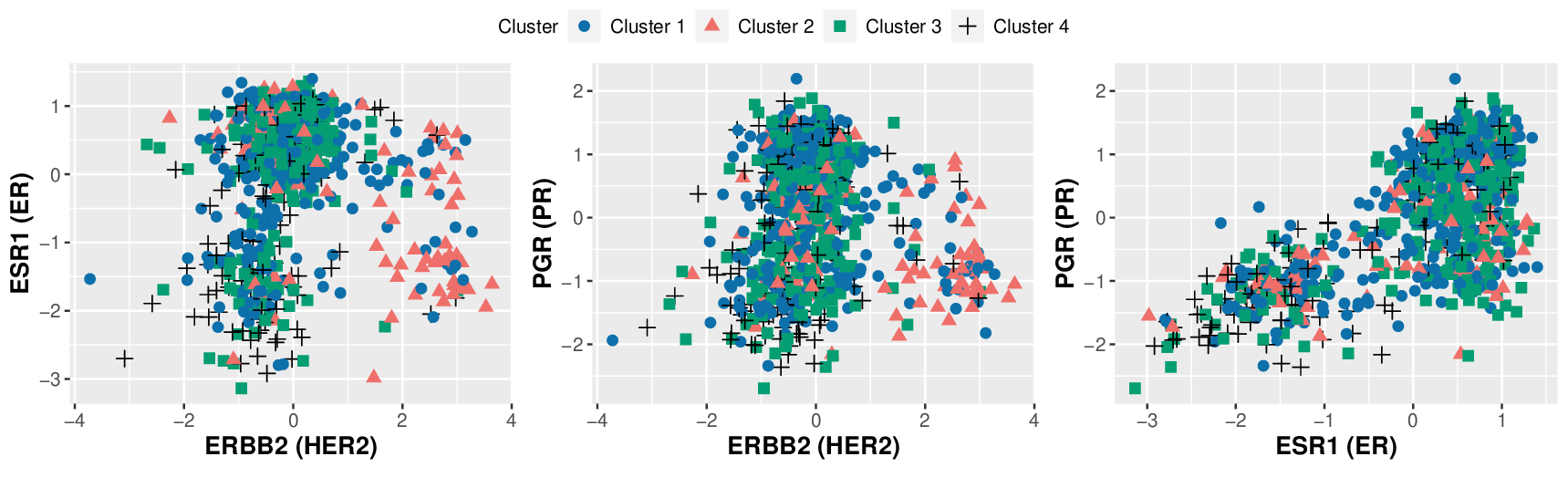}}
\caption{\small \relscale{1} Posterior estimates of cluster allocation of the graph-only model shown on pairwise scatter plots of 3 mRNA covariates: ERBB2, ESR1 and PGR.} \label{fig:real_2}
\end{figure}

We apply the PxG model with pseudo-likelihood approach and generate 10,000 MCMC samples with 1,000 burn-in samples. For comparison, the graph-only model is applied, which only uses protein information. Figure \ref{fig:real_1} shows the posterior estimates of cluster allocation under the PxG model on pairwise scatter plots of 3 mRNA covariates, and Figure \ref{fig:real_2} shows the posterior estimates of cluster allocation for the graph-only model. Under PxG, there is a clear separation among 3 of the 4 posterior clusters which can be observed from the first two scatter plots in Figure \ref{fig:real_1}, whereas the same separation is not observed under the graph-only results on Figure \ref{fig:real_2}. This suggests that the covariates influence but do not dominate the clusters in the PxG model.

Additionally, we use the total sum of squares (TSS) of covariates to measure the overall variance under a cluster allocation. It is defined as
    $\mbox{TSS} = \sum_{j=1}^{\hat{K}}\sum_{i\in \hat{S}_j}\|\mathrm{x}_i-\hat{\bm{\mu}}_j \|^2$,
where $\hat{\bm{\mu}}_j$ is the sample covariate mean of the $j$th cluster, $\hat{S}_j$ is the index set of the estimated cluster $j$, and $\hat{K}$ is the estimated number of clusters. The TSS for the PxG and graph-only models are 509,029 and 645,585, respectively, which again indicates that the covariate information  assists the clustering results for heterogeneous GGMs under PxG. The  DIC values (37,531 for the PxG model and 40,798 for the graph-only model) confirm this conclusion as well. 

\begin{figure}[H] 
\centering
\scalebox{1}{\includegraphics[width=17cm]{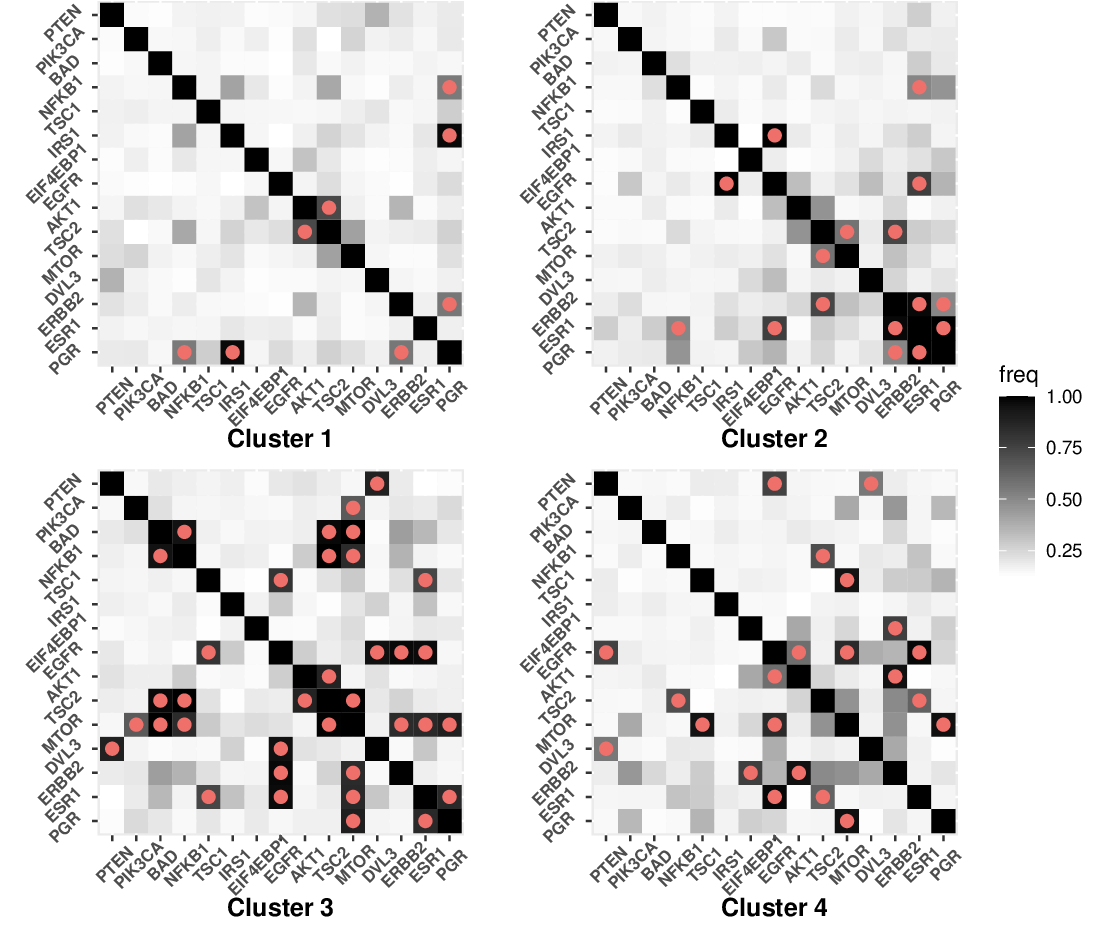}}
\caption{\small \relscale{1} Posterior edge inclusion probability for each cluster under the PxG model. Red dots indicate posterior graph selection.} \label{fig:real_3}
\end{figure}

For protein network estimation, we plot the posterior edge inclusion probability for each cluster under the PxG model in Figure \ref{fig:real_3}, where red dots indicate the included edges (inclusion probability greater than 0.5). According to the posterior estimate of the cluster allocation of the PxG model in Figure \ref{fig:real_1}, ERBB2 is highly expressed (relatively large values) in cluster 1 and it is suppressed (relatively small values) in cluster 2; ESR1 is highly expressed in both cluster 3 and 4. We observe that when ESR1 is highly expressed, the connectivity among proteins intensifies based on the protein networks from cluster 3 and 4 in Figure \ref{fig:real_3}. The connection between protein expressions of ERBB2 and PGR seems independent of the mRNA gene expression of ERBB2 since the edge between them presents in both cluster 1 and 2. Other than few exceptions, most edges among clusters do not overlap. Since cluster 3 and 4 have a similar mRNA profiles among ERBB2, ESR1 and PGR, they share the most edges than other pairwise edge comparison among clusters. The results in Figure \ref{fig:real_3} illustrate how protein connectivity change with certain mRNAs.

\section{Discussion} \label{sec:discussion}
We have introduced a novel Bayesian model, PxG, for heterogeneous graphical models with covariates. We have focused on undirected Gaussian graphical models with two different approaches, i.e., the Gaussian-G-Wishart approach and the pseudo-likelihood approach. The modularity of the proposed PxG model allows straightforward extension to other graphical models such as directed acyclic graphs. Despite the discreteness of the random partition, we have shown both theoretically and empirically that PxG is capable of recovering graphs that  continuously change with covariates. One of the important components of the proposed PxG is the PPMx model.  Recently \cite{page2021discovering} uses PPMx in regression problems to discover the interaction effects of predictors, which  are mainly categorical, on an univariate response. Note that our approach utilizes PPMx for a completely different purpose, i.e., we focus on learning the precision matrix and graph structure (conditional independencies) of a set of multivariate $Y$-variables assisted by a set of covariates $X$. Therefore, the dependencies considered in PxG and \cite{page2021discovering} are fundamentally different. Future direction of this work includes investigating other nonparametric Bayesian methods for covariate-dependent random partition models as well as a more direct approach in modeling the continuous relationship between graph/precision matrix and covariates. 

\appendix
\section{Proof of Lemma \ref{lemma}} \label{sec:lemma}

\begin{proof}
    For any $\epsilon>0$, let $K_\epsilon = \big(\frac{q^2H_c(b-a)^{\nu+2}}{\epsilon^2}\big)^{1/\nu}$, where $H_c = \max_{i,j}\{ \|\sigma^0_{ij}\|_{\mathcal{H}^\nu} \}$. Choose $a_k = a+\frac{b-a}{K_\epsilon}(k-1)$, then $a_0=a$ and $a_{K_\epsilon+1}=b$. Let $\Delta_k=(a_k,a_{k+1}]$.

    Given any $i,j\in \{1,\ldots, q\}$, for $\sigma^0_{ij}(x)$, we first choose a piecewise linear function $\sigma_{ij}(x) = \sum_{k=1}^{K_\epsilon}\delta^k_{ij}\ind_{\Delta_k}(x)$ such that
    \small
    \begin{equation*}
        |\sigma^0_{ij}(x) - \sigma_{ij}(x)| \leq \| \sigma^0_{ij}\|_{\mathcal{H}^\nu}\frac{b-a}{K_\epsilon}, \quad \text{for any } x\in(a,b]. 
    \end{equation*}
    \normalsize 
    To show this, set $\delta^k_{ij}=\sigma^0_{ij}(a_k)$. From the $\nu$-H\"{o}lder continuity,
    \small
    \begin{equation*}
        |\sigma^0_{ij}(x)-\sigma^0_{ij}(x')| \leq \| \sigma^0_{ij}\|_{\mathcal{H}^\nu} \cdot |x-x'|^\nu, \quad \text{for any } x,x'\in (a,b].
    \end{equation*}
    \normalsize
    Then, for $x\in\Delta_k$, $k=1,\ldots,K_\epsilon$, we have
    \begin{equation*}
        |\sigma^0_{ij}(x)-\sigma_{ij}(x)| = |\sigma^0_{ij}(x)-\delta^k_{ij}| = |\sigma^0_{ij}(x)-\sigma^0_{ij}(a_k)| \leq  \| \sigma^0_{ij}\|_{\mathcal{H}^\nu} \frac{(b-a)^\nu}{K_\epsilon^\nu}.
    \end{equation*}
    \normalsize
    Choose $\sigma^{0k}_{ij} = \delta^k_{ij}$. Therefore,
    \small
    \begin{align*}
        \Big\|\Omega_0^{-1}(\cdot) - \sum_{j=1}^{K_{\epsilon}}\ind_{(a_j,a_{j+1}]}(\cdot)\Omega_{0j}^{-1} \Big\|_1 & = \int_a^b \Big\|\Omega_0^{-1}(x) - \sum_{j=1}^{K_{\epsilon}}\ind_{(a_j,a_{j+1}]}(x)\Omega_{0j}^{-1} \Big\| dx \\
        & = (b-a)\cdot\mathbb{E}_x \Big\|\Omega_0^{-1}(x) - \sum_{j=1}^{K_{\epsilon}}\ind_{(a_j,a_{j+1}]}(x)\Omega_{0j}^{-1} \Big\| \\
        & \leq (b-a)\sqrt{ \mathbb{E}_x \Big\|\Omega_0^{-1}(x) - \sum_{j=1}^{K_{\epsilon}}\ind_{(a_j,a_{j+1}]}(x)\Omega_{0j}^{-1} \Big\|^2} \\
        & = \sqrt{b-a}\sqrt{\int_a^b \Big\|\Omega_0^{-1}(x) - \sum_{j=1}^{K_{\epsilon}}\ind_{(a_j,a_{j+1}]}(x)\Omega_{0j}^{-1} \Big\|^2 dx} \\
        & = \sqrt{b-a}\sqrt{ \sum_{i=1}^q\sum_{j=1}^q \int_a^b \Big|\sigma^0_{ij}(x)-\sum_{k=1}^{K_\epsilon}\ind_{\Delta_k}(x)\sigma^{0k}_{ij} \Big|^2 dx} \\
        & = \sqrt{b-a} \sqrt{\sum_{i=1}^q\sum_{j=1}^q \int_a^b |\sigma_{ij}^0(x)-\sigma_{ij}(x)| dx} \\
        & \leq \sqrt{\frac{q^2H_c(b-a)^{\nu+2}}{K_\epsilon^\nu}} = \epsilon.
    \end{align*}
    \normalsize
\end{proof}

\section{Proof of Theorem \ref{prop:suppPxG}} \label{sec:proof}
\begin{proof}
Since all entries of the true covariance matrix $\Omega_0^{-1}(x)$ are in $\mathcal{H}^\nu$, by Lemma \ref{lemma}, for a given $\epsilon>0$, there exists $K>0$, a set of positive definite matrices $\{\Omega^{-1}_{0j}\}_{j=1}^{K}$ and $\{a_j\}_{j=1}^{K}$ where $a_1=a$, $a_{K+1}=b$, such that 
\begin{equation} \label{part1}
  \Big\|\Omega^{-1}(\cdot) - \sum_{j=1}^{K}\indi_{(a_j,a_{j+1}]}(\cdot)\Omega_{0j}^{-1} \Big\|_1 <\frac{\epsilon}{3}. 
\end{equation}
Since the prior $\Omega_{j}^{-1}$ is absolutely continuous,  we have
\begin{equation*}
  \Pi\Big\{ \|\Omega_j^{-1}-\Omega_{0j}^{-1}\|< \frac{\epsilon}{3K} \Big\} >0, \text{ for all } j = 1, \ldots, K.
\end{equation*}
Hence,
\begin{equation} \label{part2}
  \Pi \Big\{ \sum_{j=1}^{K}\|\Omega_j^{-1}-\Omega_{0j}^{-1}\|< \frac{\epsilon}{3} \Big\} >0.
\end{equation}
Next, we shall carefully choose neighborhoods $B_{\mu_{j}}$ and $B_{\sigma^2_j}$ so that $\pi_j(x)$ approximates an indicator function. Without loss of generality, assume intervals $(a_j,a_{j+1}]$, $j=1,\ldots, K$ have equal length $L=(b-a)/K$. We subdivide $(a_j,a_{j+1}]$ into $K'$ equal length intervals $(a_j,a_{j+1}] = \cup_{l=1}^{K'}(a_{jl}, a_{j(l+1)}]$ for $j=1,\ldots, K$ and the length of $(a_{jl},a_{j(l+1)}]$ is $\frac{L}{K'}$.
Define
\begin{align*}
    \pi_{jl}(x) &= \frac{\pi_{jl}\frac{1}{\sqrt{\sigma_{jl}^2}}\exp\Big\{-\frac{(x-\mu_{jl})^2}{2\sigma_{jl}^2} \Big\}}
    {\sum_{t=1}^K\sum_{s=1}^{K'}\pi_{ts}\frac{1}{\sqrt{\sigma_{ts}^2}}\exp\Big\{-\frac{(x-\mu_{ts})^2}{2\sigma_{ts}^2} \Big\}}, \,\,j=1,\ldots,K,\,\, l=1,\dots,K'.
\end{align*}
Let $\sigma^2_{jl}\equiv \sigma^2$ and $\pi_{jl}\equiv \pi$ for all $j$ and $l$, then $\pi_{jl}(x)$ can be simplified to
\begin{align*}
    \pi_{jl}(x) &= \frac{\exp\Big\{-\frac{(x-\mu_{jl})^2}{2\sigma_{jl}^2} \Big\}}
    {\sum_{t=1}^K\sum_{s=1}^{K'}\exp\Big\{-\frac{(x-\mu_{ts})^2}{2\sigma_{ts}^2} \Big\}}, \,\,j=1,\ldots,K,\,\, l=1,\dots,K',
\end{align*}
where $\mu_{jl}$ and $\sigma^2$ are defined as follows.
\begin{equation*}
  \mu_{jl} \in 
  \Big[\frac{a_{jl}+a_{j(l+1)}}{2}-\frac{L}{6{K'}}, \frac{a_{jl}+a_{j(l+1)}}{2}+\frac{L}{6{K'}}\Big],\qquad
  \sigma^2 \in 
  \bigg[\frac{L^2}{2{K'}^3}, \frac{L^2}{{K'}^3}\bigg].
\end{equation*}
With  absolutely continuous priors on $\mu_{jl}$ and $\sigma^2$, the prior probabilities of $B_{\mu_{jl}}$ and $B_{\sigma^2_{jl}}$ are non-zero. Let $M = \max\{\|\Omega_{0j}^{-1}\|\}_{j=1}^{K}$. For any $j = 1, \ldots, K$, $l=1,\dots,K'$ and the $\epsilon$ above, choose
\begin{equation*}
    K' = \max\Bigg\{ \frac{12LM}{5\epsilon}, 30\log\Big(\frac{12LMK}{\epsilon}\Big) \Bigg\} + 1.    
\end{equation*}
In the following, we sub-divide $\big\|\pi_{jl}(x)-\mathbb{I}_{(a_{jl},a_{j(l+1)}]}(x)\big\|_1$ into four mutually exclusive integrals as follows.
\begin{align*}
    \Big\|\pi_{jl}(x)-\mathbb{I}_{(a_{jl},a_{j(l+1)}]}(x)\Big\|_1
    & = \int_a^b \Big|\pi_{jl}(x)-\mathbb{I}_{(a_{jl},a_{j(l+1)}]}(x)\Big| dx \\
    & = \int_{x\in\big(a_{jl}, a_{jl}+\frac{L}{5K'}\big]}\Big|\pi_{jl}(x)-1\Big|dx  +  \int_{x\in\big(a_{jl}+\frac{L}{5K'}, a_{j(l+1)}-\frac{L}{5K'}\big]}\Big|\pi_{jl}(x)-1\Big|dx + \\ & \int_{x\in\big(a_{j(l+1)}-\frac{L}{5K'}, a_{j(l+1)}\big]} \Big|\pi_{jl}(x)-1\Big|dx + \int_{x\not\in\big(a_{jl},a_{j(l+1)}\big]}\Big|\pi_{jl}(x)\Big|dx \\
    & := \mbox{I}'_{jl} + \mbox{I}''_{jl} + \mbox{I}'''_{jl} + \mbox{I}''''_{jl}.
\end{align*}
First, we have
\begin{equation*}
    \mbox{I}'_{jl} \leq \frac{L}{5K'}\cdot 1 < \frac{\epsilon}{12\|\Omega^{-1}_{0j}\|}, \qquad \mbox{I}'''_{jl} \leq \frac{L}{5K'}\cdot 1 < \frac{\epsilon}{12\|\Omega^{-1}_{0j}\|}.
\end{equation*}
Moreover,
\begin{align*}
    \mbox{I}''_{jl} & = \int_{x\in\big(a_{jl}+\frac{L}{5K'}, a_{j(l+1)}-\frac{L}{5K'}\big]}
    \Bigg| \frac{1} {1 + \sum\sum_{(t,s)\neq (j,l)}  \exp\Big\{-\frac{(x-\mu_{ts})^2}{2\sigma^2} + \frac{(x-\mu_{jl})^2}{2\sigma^2} \Big\} } -1 \Bigg| dx \\
    & \leq \int_{x\in\big(a_{jl}+\frac{L}{5K'}, a_{j(l+1)}-\frac{L}{5K'}\big]}
    {\sum\sum}_{(t,s)\neq (j,l)} \exp\Big\{-\frac{(x-\mu_{ts})^2}{2\sigma^2} +  \frac{(x-\mu_{jl})^2}{2\sigma^2} \Big\} dx \\
    & \leq \frac{L}{K'}\cdot KK' \exp\Bigg\{-\frac{\Big(\frac{L}{2K'}-\frac{L}{6K'}+\frac{L}{5K'} \Big)^2}{2\frac{L^2}{{K'}^3}} + \frac{\Big(\frac{L}{2K'}-\frac{L}{5K'}+\frac{L}{6K'}\Big)^2}{2\frac{L^2}{{K'}^3}} \Bigg\} \\
  & \leq LK \exp \Big\{-\frac{K'}{30} \Big\} < \frac{\epsilon}{12\|\Omega_{0j}^{-1}\|},
\end{align*}
and
\begin{align*}
    \mbox{I}''''_{jl} & = \int_{x\not\in\big(a_{jl},a_{j(l+1)}\big]}  \Bigg| \frac{1} {1 + \sum\sum_{(t,s)\neq (j,l)}  \exp\Big\{-\frac{(x-\mu_{ts})^2}{2\sigma^2} + \frac{(x-\mu_{jl})^2}{2\sigma^2} \Big\} } \Bigg| dx \\ 
    & \leq \frac{L}{K'} \exp\Big\{\frac{(x-\mu_{j(l+2)})^2}{2\sigma^2} - \frac{(x-\mu_{jl})^2}{2\sigma^2} \Big\}\Bigg|_{x=a{j(l+2)}} \\
    & \leq L \exp\Bigg\{ \frac{\Big(\frac{L}{2K'}+\frac{L}{6K'} \Big)^2}{2\frac{L^2}{{K'}^3}} - \frac{\Big(\frac{3L}{2K'} - \frac{L}{6K'} \Big)^2}{2\frac{L^2}{{K'}^3}}  \Bigg\} \\
    & \leq L \exp\Big\{ -\frac{2}{3}K'  \Big\} < \frac{\epsilon}{12\|\Omega_{0j}^{-1}\|}.
\end{align*}

\noindent Combining the bounds for $\mbox{I}'_{jl}$, $\mbox{I}''_{jl}$, $\mbox{I}'''_{jl}$ and $\mbox{I}''''_{jl}$, we have
\begin{align*}
  \Pi\Big\{ \Big\| \ind_{(a_{jl},a_{j(l+1)}]}(\cdot)\Omega_{0j}^{-1} - \pi_{jl}(\cdot)\Omega_{0j}^{-1}  \Big\|_1 <\frac{\epsilon}{3} \mid \pi \Big\}>0, \,\, \text{for all } j =1,\ldots, K, l=1,\ldots,K'.
\end{align*}
Therefore, 
\begin{equation} \label{part3}
    \Pi\Big\{ \Big\| \sum_{j=1}^{K}\sum_{l=1}^{K'} \indi_{(a_{jl},a_{j(l+1)}]}(\cdot)\Omega_{0j}^{-1} - \sum_{j=1}^{K}\sum_{l=1}^{K'}\pi_{jl}(\cdot)\Omega_{0j}^{-1}  \Big\|_1 <\frac{\epsilon}{3} \mid \bm{\pi} \Big\}>0.
\end{equation} 
To finalize, observe by repeated application of triangle inequality,  
\begin{align*}
  &\phantom{\,\,\,\,\,\,\,\,} \Big\| \Omega^{-1}(\cdot) - \sum_{j=1}^K\sum_{l=1}^{K'}\pi_{jl}(\cdot)\Omega_j^{-1} \Big\|_1 \\ 
    & \leq \Big\| \Omega^{-1}(\cdot) - \sum_{j=1}^K\sum_{l=1}^{K'} \indi_{(a_{jl},a_{j(l+1)}]}(\cdot)\Omega_{0j}^{-1}\Big\|_1 + \Big\|\sum_{j=1}^K\sum_{l=1}^{K'} \indi_{(a_{jl},a_{j(l+1)}]}(\cdot)\Omega_{0j}^{-1} - \sum_{j=1}^K\sum_{l=1}^{K'}\pi_{jl}(\cdot)\Omega_{0j}^{-1} \Big\|_1 + \\ & \quad \, \, \Big\|\sum_{j=1}^K\sum_{l=1}^{K'} \pi_{jl}(\cdot)\Omega_{0j}^{-1} - \sum_{j=1}^K\sum_{l=1}^{K'}\pi_{jl}(\cdot)\Omega_{j}^{-1} \Big\|_1 \\
  & \leq \frac{\epsilon}{3} + \frac{\epsilon}{3} + \sum_{j=1}^{K}\sum_{l=1}^{K'}\| \Omega_{j}^{-1}-\Omega_{0j}^{-1} \| < \epsilon.
\end{align*}
Apply (\ref{part1}), (\ref{part2}) and (\ref{part3}),
\begin{equation*}
    \Pi\Big \{ \big\| \Omega^{-1}(\cdot) - \sum_{j=1}^{K}\pi_j(\cdot)\Omega_{j}^{-1} \big\|_1 < \epsilon  \mid \bm{\pi} \Big \} > 0.
\end{equation*}
By integrating across the prior density of $\pi$, we finally conclude the proof. 
\end{proof}

\section{Proof of Theorem \ref{th:contraction} } \label{proof2}
To prove Theorem \ref{th:contraction}, we simply need to show the prior mass condition holds. To that end, first define the Kullback-Leibler neighborhood of $p_{\Theta_0}$ with radius $\epsilon_n$ as
    \begin{equation*}
        B_n(p_{\Theta_0}, \epsilon_n) = \Big\{\Theta\in \mathscr{X}: \int p_{\Theta_0}\log\Big(\frac{p_{\Theta_0}}{p_{\Theta}} \Big) d\mu \leq \epsilon_n^2, \int p_{\Theta_0}\log^2\Big(\frac{p_{\Theta_0}}{p_{\Theta}} \Big) d\mu \leq \epsilon_n^2 \Big\},
    \end{equation*}
where $\mathscr{X}$ is the parameter space, $\mu$ is a dominating measure on the common probability space.

\begin{theorem} {\normalfont(Prior Mass Condition).} Suppose Assumption \ref{assump:true} and \ref{assump:prior} hold. Then, for any $C>0$ and all sufficiently large $n$,
    \begin{equation*}
        \Pi\{B_n(p_{\Theta_0},\epsilon_n)\}\geq \exp(-C n\epsilon^2_n ),
    \end{equation*}
where $\epsilon_n=\sqrt{\frac{K(p+q)\log n}{n}}$. Here $\Pi(\cdot)$ is the joint prior probability of all parameters.
\end{theorem}

\begin{proof}
The structure of this proof is similar to the proof of Theorem 3.1 in \citet{norets2017adaptive}, with a few important differences as we shall see below. All constants in the proof are independent of $p,q,n,K$. Let $\bm{\mu}_{0j}=(\mu_{0j1},\ldots,\mu_{0jp})$, $\bm{\mu}_j=(\mu_{j1},\ldots,\mu_{jp})$, $\Sigma^x_j=\mbox{diag}(\sigma^2_{j1},\ldots,\sigma^2_{jp})$. Denote $0<\lambda_{j1}\leq\ldots\leq\lambda_{jq}$ to be the eigenvalues of $\Sigma_{G_j}$. 
Let $\delta_n=\epsilon_n/\Delta$, where $\Delta=\max\big\{\log(Ke^{-C^*_6(p+q)}), p(e^{-C^*_1q}+e^{-C^*_2p})(\log n + p + q) \big\}$ for $\Theta$ in set
\begin{align*}
    S_{\Theta_0}  = & \Big\{ (\pi_j, \Sigma_{G_j},\bm{\mu}_j,\Sigma^x_j),\,\,
    \sum_{j=1}^K |\pi_{0j}-\pi_j|\leq 2\delta_n^2,\,\,
    \min_{j=1,\ldots,K}\pi_j\geq \delta_n^4/2,
    \frac{|\mu_{ji}-\mu_{0ji}|}{\sigma_{0ji}} \leq \delta^2_n,\,\,
    \sigma^2_{ji} \in \Big[\frac{\sigma^2_{0ji}}{1+\delta_n^2} ,\sigma^2_{0ji}\Big], \\
    & 1 \leq \frac{\lambda_{jk}}{\lambda_{0jk}} \leq 1+\frac{\delta_n^4}{q}, \,\,
    i = 1,\ldots, p,\,\, k=1,\ldots, q,\,\, j=1,\dots,K \Big\},
\end{align*}
we have
\small
\begin{align*}
    & d^2_H\Big\{p(\cdot \mid \Theta_0), p(\cdot \mid \Theta)\Big\} 
    \leq \Big\|\sum_{j=1}^K \pi_{0j} \cdot \phi_{\Sigma^0_{G_{0j}}} \cdot \phi_{\bm{\mu}_{0j},\Sigma^x_{0j}}  -  \sum_{j=1}^K \pi_j \cdot \phi_{\Sigma_{G_j}} \cdot \phi_{\bm{\mu}_j,\Sigma^x_j} \Big\|_1 \\
    & \leq \Big\|\sum_{j=1}^K \pi_{0j}  \phi_{\Sigma^0_{G_{0j}}}  \phi_{\bm{\mu}_{0j},\Sigma^x_{0j}} - \sum_{j=1}^K \pi_{0j}  \phi_{\Sigma_{G_j}}  \phi_{\bm{\mu}_j,\Sigma^x_j} + \sum_{j=1}^K \pi_{0j}  \phi_{\Sigma_{G_j}}  \phi_{\bm{\mu}_j,\Sigma^x_j} - \sum_{j=1}^K \pi_j  \phi_{\Sigma_{G_j}}  \phi_{\bm{\mu}_j,\Sigma^x_j} \Big\|_1\\
    & \leq \sum_{j=1}^K |\pi_{0j}-\pi_j| + \sum_{j=1}^K\pi_{0j}\Bigg[
    \Big\|\phi_{\Sigma^0_{G_{0j}}}  \phi_{\bm{\mu}_{0j},\Sigma^x_{0j}} - \phi_{\Sigma^0_{G_{0j}}}  \phi_{\bm{\mu}_{j},\Sigma^x_j}\Big\|_1 + \Big\| \phi_{\Sigma^0_{G_{0j}}}  \phi_{\bm{\mu}_j,\Sigma^x_j} - \phi_{\Sigma_{G_j}}  \phi_{\bm{\mu}_{j},\Sigma^x_{j}} \Big\|_1 \Bigg] \\
    & \leq \sum_{j=1}^K |\pi_{0j}-\pi_j| + \sum_{j=1}^K \pi_{0j} (2\pi)^{-\frac{q}{2}}|\Sigma^0_{G{0j}}|^{-\frac{1}{2}} \Big\|\phi_{\bm{\mu}_{0j},\Sigma^x_{0j}} - \phi_{\bm{\mu}_{j},\Sigma^x_j}\Big\|_1 + \sum_{j=1}^K \pi_{0j}(2\pi)^{-\frac{p}{2}}|\Sigma^x_j|^{-\frac{1}{2}} \Big\| \phi_{\Sigma^0_{G_{0j}}} - \phi_{\Sigma_{G_j}} \Big\|_1 \\
    & \leq \sum_{j=1}^K |\pi_{0j}-\pi_j| + \sum_{j=1}^K \pi_{0j} |\Sigma^0_{G_{0j}}|^{-\frac{1}{2}} \Big\|\phi_{\bm{\mu}_{0j},\Sigma^x_{0j}} - \phi_{\bm{\mu}_{j},\Sigma^x_j}\Big\|_1 + \sum_{j=1}^K \pi_{0j}|\Sigma^x_j|^{-\frac{1}{2}} \Big\| \phi_{\Sigma^0_{G_{0j}}} - \phi_{\Sigma_{G_j}} \Big\|_1.
\end{align*}
\normalsize
For $j=1,\ldots,K$, by Proposition \ref{normaltv},
\begin{align*}
    \pi_{0j} |\Sigma^0_{G_{0j}}|^{-\frac{1}{2}} \Big\|\phi_{\bm{\mu}_{0j},\Sigma^x_{0j}} - \phi_{\bm{\mu}_{j},\Sigma^x_j}\Big\|_1
    & \leq \pi_{0j} \Big(\prod_{i=1}^q \lambda^0_{ji}\Big)^{-\frac{1}{2}}\Bigg\{\sum_{i=1}^p\Bigg(\frac{\sigma^2_{ji}}{\sigma^2_{0ji}}-1 - \log\frac{\sigma^2_{ji}}{\sigma^2_{0ji}}\Bigg) + \sum_{i=1}^p\frac{(\mu_{0ji}-\mu_{ji})^2}{\sigma^2_{0ji}} \Bigg\}^{\frac{1}{2}} \\
    & \leq \pi_{0j} \Big(\prod_{i=1}^q \lambda^0_{ji}\Big)^{-\frac{1}{2}} \sum_{i=1}^p \Bigg(\frac{|\mu_{0ji}-\mu_{ji}|}{\sigma_{0ji}} + \Bigg|\frac{\sigma^2_{ji}}{\sigma^2_{0ji}}-1 - \log\frac{\sigma^2_{ji}}{\sigma^2_{0ji}}\Bigg|^{\frac{1}{2}} \Bigg) \\
    & \leq \pi_{0j} \Big(\prod_{i=1}^q \lambda^0_{ji}\Big)^{-\frac{1}{2}} \sum_{i=1}^p \Bigg(\frac{|\mu_{0ji}-\mu_{ji}|}{\sigma_{0ji}} + C_1\Bigg|\frac{\sigma^2_{ji}}{\sigma^2_{0ji}}-1 \Bigg| \Bigg) \\
    & \leq \pi_{0j} (\lambda^0_m)^{-\frac{q}{2}} \cdot p(1+C_1)\delta_n^2,
\end{align*}
where the second to the last step follows $|\log x-x+1|\leq C_1|x-1|^2$ for $x$ in a neighborhood of 1, where $C_1>0$ is a constant. Next, by following the proof from \citet{shen2013adaptive}, we show that
\begin{align*}
    \pi_{0j}|\Sigma^x_j|^{-\frac{1}{2}} \Big\| \phi_{\Sigma^0_{G_{0j}}} - \phi_{\Sigma_{G_j}} \Big\|_1
    & \leq \pi_{0j} \Big(\prod_{i=1}^p \sigma^2_{ji} \Big)^{-\frac{1}{2}} \Bigg(\Big\| \phi_{\Sigma^0_{G_{0j}}} - \phi_{\Sigma'_j} \Big\|_1 + \Big\| \phi_{\Sigma'_j} - \phi_{\Sigma_{G_j}} \Big\|_1\Bigg) \\
    & \leq \pi_{0j} \Big(\frac{\sigma^2_m}{2}\Big)^{-\frac{p}{2}}\Bigg(\Big\| \phi_{\Sigma^0_{G_{0j}}} - \phi_{\Sigma'_j} \Big\|_1 + \Big\| \phi_{\Sigma'_j} - \phi_{\Sigma_{G_j}} \Big\|_1\Bigg) \leq 2 \pi_{0j} \Big(\frac{\sigma^2_m}{2}\Big)^{-\frac{p}{2}}\delta_n^2,
\end{align*}
where $\Sigma'_j$ is a positive definite matrix constructed as follows.

Let $\|\cdot\|_2$ be the matrix spectral norm and $\|\cdot\|_{\max}$ be the maximum of the absolute value of the matrix elements. Assume $\Sigma^0_{G_{0j}}=P_{0j}\Lambda^*_{0j}P_{0j}^T$ and find the orthogonal matrix $P_j$ such that $\Sigma_{G_j} = P_j\Lambda^*_j P_j^T$ and $\|P_{0j}-P_j \|_2\leq \frac{\delta_n^4}{3q\lambda^0_M/\lambda^0_m}$, where $P_{0j}P_{0j}^T=I$ and $P_jP_j^T=I$. Define $\Sigma'_j=P_j\Lambda^*_{0j} P_j^T$ and $Q_j =P_{0j}^TP_j$.
\vspace{-0.5cm}
\begin{align*}
    & \Big\|\phi_{\Sigma_{G_j}} - \phi_{\Sigma'_j} \Big\|_1 
    \leq \big\{ \mbox{tr}(\Sigma'^{-1}_j\Sigma_{G_j}) -\log |\Sigma'^{-1}_j\Sigma_{G_j}| -q \big\}^{\frac{1}{2}} \\
    & = \Bigg(\sum_{i=1}^q\frac{\lambda_{ji}}{\lambda_{0ji}} - \log\prod_{j=1}^q\frac{\lambda_{ji}}{\lambda_{0ji}}-q \Bigg)^{\frac{1}{2}} \\
    & \leq (q + \delta_n^4 - 0 - q )^{\frac{1}{2}} = \delta_n^2, \\
    & \Big\|\phi_{\Sigma'_j} - \phi_{\Sigma^0_{G_0}} \Big\|_1 
    \leq \big\{ \mbox{tr}({\Sigma^0_{G_0}}^{-1}\Sigma'_j) -\log |{\Sigma^0_{G_0}}^{-1}\Sigma'_j| -q \big\}^{\frac{1}{2}} \\
    & = \big\{\mbox{tr}(P_{0j}{\Lambda^*_{0j}}^{-1}P^T_{0j}P_j\Lambda^*_{0j}P_j^T)-q \Big\}^{\frac{1}{2}} \\
    & = \big\{\mbox{tr}(Q_j\Lambda^*_{0j}Q^T_j{\Lambda^*_{0j}}^{-1} - I ) \big\}^{\frac{1}{2}}.
\end{align*}
Let $Q_j = I + B_j$, $\|B_j\|_{\max} \leq \|B_j\|_2 = \| P_{0j}^TP_j-I\|_2 = \|P_{0j}-P_j\|_2 \leq \frac{\delta_n^4}{3q\lambda^0_M/\lambda^0_m}$.
\begin{align*}
    \mbox{tr}(Q_j\Lambda^*_{0j}Q_j{\Lambda^*_{0j}}^{-1}- I ) 
    = \mbox{tr}(B_j + \Lambda^*_{0j}B_j^T{\Lambda^*_{0j}}^{-1} + B_j\Lambda^*_{0j}B_j^T{\Lambda^*_{0j}}^{-1})
    \leq 3q\|B_j\|_{\max}\frac{\lambda^0_M}{\lambda^0_m} \leq \delta_n^4.
\end{align*}
Therefore, 
$$
d^2_H\Big\{p(\cdot \mid \Theta_0), p(\cdot \mid \Theta)\Big\}  \leq \Bigg\{2 + p(1+C_1)(\lambda^0_m)^{-\frac{q}{2}} + 2\Big(\frac{\sigma^2_m}{2}\Big)^{-\frac{p}{2}} \Bigg\} \delta_n^2 
\leq p(e^{-C^*_1q}+e^{-C^*_2p})\delta_n^2
:=C_2\delta_n^2.
$$

For any $\Theta\in S_{\Theta_0}$, consider a lower bound on the ratio $p_{\Theta}/p_{\Theta_0}=p(\mathrm{y},\mathrm{x}\mid\Theta)/p(\mathrm{y},\mathrm{x}\mid \Theta_0)$. Let $a_n=2A_1\sqrt{\log n}+A_2$, where $A_1>0$ and $A_2>0$ are two constants depending on $\Theta_0$,
then
$$
    \mathbb{P}_{\Theta_0}(\|(\mathrm{y},\mathrm{x})\|>a_n) = \sum_{j=1}^K \pi_j \mathbb{P}_{\Theta_0}(\|(\mathrm{y},\mathrm{x})\|>a_n\mid z = j) 
    \leq \frac{1}{n^4} \leq \frac{\delta_n^4}{\log^4(\delta_n^5)},
$$
where $z$ indicates which cluster $(\mathrm{y},\mathrm{x})$ is in.

For $\|(\mathrm{x},\mathrm{y})\| \leq a_n$, where $\|\cdot\|$ is the Euclidean norm, for all sufficiently large $n$,
\small
\begin{align*}
  & p(\mathrm{y},\mathrm{x}\mid \Theta)
    = \sum_{j=1}^K \pi_j \cdot (2\pi)^{-\frac{p}{2}}|\Sigma^x_j|^{-\frac{1}{2}}\exp\Bigg\{-\sum_{i=1}^p\frac{(x_i-\mu_{ji})^2}{2\sigma_{ji}^2}\Bigg \} \cdot (2\pi)^{-\frac{q}{2}}|\Sigma_{G_j}|^{-\frac{1}{2}}\exp\Bigg\{-\frac{1}{2}\mathrm{y}^T\Sigma_{G_j}^{-1}\mathrm{y} \Bigg\} \\
    & \geq \min_j\pi_j \cdot (2\pi)^{-\frac{p+q}{2}}(\sigma_M^2)^{-\frac{p}{2}}\lambda_M^{-\frac{q}{2}}\Bigg(1+\frac{\epsilon_n}{q}\Bigg)^{-\frac{q}{2}} \cdot \sum_{j=1}^K \exp\Bigg\{-\frac{\|\mathrm{x}-\bm{\mu}_j \|^2}{2\sigma_m^2}\Bigg \} \exp\Bigg\{ -\frac{1}{2}\|\Sigma_{G_j}^{-\frac{1}{2}}\mathrm{y}\|^2 \Bigg\} \\
    & \geq (\delta_n^4/2) (2\pi)^{-\frac{p+q}{2}}(\sigma_M^2)^{-\frac{p}{2}}\lambda_M^{-\frac{q}{2}}\Bigg(1+\frac{\epsilon_n}{q}\Bigg)^{-\frac{q}{2}}
    \sum_{j=1}^K \exp\bigg\{ -\frac{\sum_{i=1}^p(\epsilon_n\sigma_M+|\mu_{0ji}|)^2}{2\sigma^2_m} \bigg\}
    \exp\bigg\{-\frac{1}{2}\bigg(\frac{\|\mathrm{x}\|^2}{\sigma^2_m} + \lambda_M\|\mathrm{y}\|^2 \bigg) \bigg\} \\
    & \geq e^{-C^*_3(p+q)}K\delta_n^4\exp(-C^*_4a_n^2) = C_3K\delta^4_n \exp(-C_4a_n^2),
\end{align*}
\normalfont
and
\begin{equation*}
  \frac{p(\mathrm{y},\mathrm{x}\mid\Theta)}{p(\mathrm{y},\mathrm{x}\mid\Theta_0)} \geq  e^{-C^*_5(p+q)}K\delta_n^4\exp(-C^*_4a_n^2)  :=\lambda_n, 
\end{equation*}
where $C_3,C_4,C_5>0$. \\

For $\|(\mathrm{x},\mathrm{y})\|> a_n$ and all sufficiently large $n$,
\begin{equation*}
    \frac{p(\mathrm{y},\mathrm{x}\mid\Theta)}{p(\mathrm{y},\mathrm{x}\mid \Theta_0)} 
     \geq e^{-C^*_6(p+q)}K\delta_n^4
    \exp(-C^*_7\|(\mathrm{x},\mathrm{y})\|^2) 
    \geq C_7K\delta_n^4 \exp(-C_6\|(\mathrm{x},\mathrm{y})\|^2),
\end{equation*}
where $C_6 = \max\{ 1/(2\sigma^2_m),\lambda_M/2\}$. \\

Consider all sufficiently large $n$ such that $\lambda_n\leq 1/e$. For any $\Theta\in S_{\Theta_0}$,
\begin{align*}
    & \int \Bigg(\log\frac{p(\mathrm{y},\mathrm{x}\mid\Theta_0)}{p(\mathrm{y},\mathrm{x}\mid\Theta)}\Bigg)^2\mathbbm{1}\Bigg(\frac{p(\mathrm{y},\mathrm{x}\mid\Theta)}{p(\mathrm{y},\mathrm{x}\mid\Theta_0)}\leq\lambda_n \Bigg) p(\mathrm{y},\mathrm{x}\mid\Theta_0)\, d\mathrm{x}\,d\mathrm{y} \\
    & = \int \Bigg(\log\frac{p(\mathrm{y},\mathrm{x}\mid\Theta_0)}{p(\mathrm{y},\mathrm{x}\mid\Theta)}\Bigg)^2\mathbbm{1}\Bigg(\frac{p(\mathrm{y},\mathrm{x}\mid\Theta)}{p(\mathrm{y},\mathrm{x}\mid\Theta_0)}\leq\lambda_n, \|(\mathrm{x},\mathrm{y})\|>a_n \Bigg) p(\mathrm{y},\mathrm{x}\mid\Theta_0)\, d\mathrm{x}\,d\mathrm{y} \\
    &\leq 2{C^*_7}^2\int_{\|(\mathrm{x},\mathrm{y})\|>a_n} \|(\mathrm{x},\mathrm{y})\|^4 p(\mathrm{y},\mathrm{x}\mid\Theta_0)\, d\mathrm{x}\,d\mathrm{y} + 2\log^2(e^{-C^*_6(p+q)}K\delta_n^4)\int_{\|(\mathrm{x},\mathrm{y})\|>a_n} p(\mathrm{y},\mathrm{x}\mid\Theta_0)\, d\mathrm{x}\,d\mathrm{y}  \\
    & \leq \{ {2C^*_7}^2(\mathbb{E}_0\|(\mathrm{x},\mathrm{y})\|^4)^{\frac{1}{2}} + 2\log^2(e^{-C^*_6(p+q)}K\delta_n^4) \} [\mathbb{P}_{\Theta_0}(\|\mathrm{x},\mathrm{y})\|\geq a_n)]^{\frac{1}{2}} \\
    & \leq 2{C^*_7}^2(\mathbb{E}_0\|(\mathrm{x},\mathrm{y})\|^4)^{\frac{1}{2}} \frac{\delta_n^2}{\log^2(\delta_n^5)} + 2\log^2(e^{-C^*_6(p+q)}K\delta_n^4) \frac{\delta_n^2}{\log^2(\delta_n^5)} \leq C^*_8\log^2(e^{-C^*_6(p+q)}K)\delta^2_n.
\end{align*}
Also note that
\begin{align*}
    & \log\frac{p(\mathrm{y},\mathrm{x}\mid\Theta_0)}{p(\mathrm{y},\mathrm{x}\mid\Theta)}\mathbbm{1}\Bigg(\frac{p(\mathrm{y},\mathrm{x}\mid\Theta)}{p(\mathrm{y},\mathrm{x}\mid\Theta_0)}\leq\lambda_n \Bigg) p(\mathrm{y},\mathrm{x}\mid\Theta_0)\, d\mathrm{x}\,d\mathrm{y} \\
    & \leq \Bigg(\log\frac{p(\mathrm{y},\mathrm{x}\mid\Theta_0)}{p(\mathrm{y},\mathrm{x}\mid\Theta)}\Bigg)^2\mathbbm{1}\Bigg(\frac{p(\mathrm{y},\mathrm{x}\mid\Theta)}{p(\mathrm{y},\mathrm{x}\mid\Theta_0)}\leq\lambda_n\Bigg) p(\mathrm{y},\mathrm{x}\mid\Theta_0)\, d\mathrm{x}\,d\mathrm{y},
\end{align*}
Thus,
\begin{equation*}
    \int \log\frac{p(\mathrm{y},\mathrm{x}\mid\Theta_0)}{p(\mathrm{y},\mathrm{x}\mid\Theta)}\mathbbm{1}\Bigg(\frac{p(\mathrm{y},\mathrm{x}\mid\Theta)}{p(\mathrm{y},\mathrm{x}\mid\Theta_0)}\leq\lambda_n \Bigg) p(\mathrm{y},\mathrm{x}\mid\Theta_0)\, d\mathrm{x}\,d\mathrm{y}
    \leq C^*_8\log^2(e^{-C^*_6(p+q)}K)\delta^2_n.
\end{equation*}
By Proposition \ref{KLbounds},
\begin{align*}
    \mathbb{E}[\log(p_{\Theta_0}/p_{\Theta})] & \leq 
    p(e^{-C^*_1q}+e^{-C^*_2p})\delta_n^2 \Big(1 + 2\log\frac{1}{e^{-C^*_5(p+q)}K\delta_n^4}+2C^*_4a_n^2\Big)  + 2C^*_8\log^2(e^{-C^*_6(p+q)}K)\delta^2_n \\
    & \leq C^*_9 p(e^{-C^*_1q}+e^{-C^*_2p})(\log n + p + q) \delta_n^2 + 2C^*_8\log^2(e^{-C^*_6(p+q)}K)\delta^2_n \leq A\epsilon_n^2, \\
    \mathbb{E}[\log(p_{\Theta_0}/p_{\Theta})]^2 & \leq p(e^{-C^*_1q}+e^{-C^*_2p})\delta_n^2  \Big\{12 + 2\Big(\log\frac{1}{e^{-C^*_5(p+q)}K\delta_n^4}+C^*_4a_n^2\Big)^2\Big\}  + 8C^*_8\log^2(e^{-C^*_6(p+q)}K)\delta^2_n \\
    & \leq C^*_{10} p(e^{-C^*_1q}+e^{-C^*_2p})(\log n + p + q)^2 \delta_n^2 + 8C^*_8\log^2(e^{-C^*_6(p+q)}K)\delta^2_n \leq A\epsilon_n^2.
\end{align*}

\noindent Therefore, both $\mathbb{E}[\log(p_{\Theta_0}/p_{\Theta})]$ and $\mathbb{E}[\log(p_{\Theta_0}/p_{\Theta})]^2$ are bounded by $A\epsilon_n^2$ for some constant $A$. 

Finally, we calculate a lower bound on the prior probability of $\{\Theta\in S_{\Theta_0} \}$. From Lemma 10 of \citet{ghosal2007posterior}, for some constants $C_{10}>0$ and all sufficiently large $n$,
\begin{align*}
  \Pi\Bigg( \sum_{j=1}^K |\pi_{0j}-\pi_j|\leq 2\delta_n^2,
    \min_{j=1,\ldots,K}\pi_j\geq \delta^4_n/2, \Bigg) & \geq \exp\{-C_{10}K\log(1/\delta_n) \}.
\end{align*}
Let $m_{ji}$ be the lower bound of the prior density of $\mu_{ji}$ in the interval of $[\mu_{0ji}-\sigma_{0ji}\delta_n^2, \mu_{0ji}+\sigma_{0ji}\delta_n^2]$, $j=1,\ldots,K, i = 1,\ldots, p$, then
\begin{align*}
    \Pi\Bigg( \frac{|\mu_{ji}-\mu_{0ji}|}{\sigma_{0ji}} \leq \delta_n^2, \,\, j=1,\ldots,K, i = 1,\ldots, p \Bigg) & \geq \prod_{j=1}^K\prod_{i=1}^p m_{ji}2\sigma_{0ji}\delta_n^2 \geq C_{12}\exp \{ -C_{11}Kp\log(1/\delta_n)\}.
\end{align*}
Let $m'_{ji}$ be the lower bound of the prior density of $\sigma_{ji}$ in the interval of $\Big[\frac{\sigma^2_{0ji}}{1+\delta_n^2} ,\sigma^2_{0ji}\Big]$, $j=1,\ldots,K, i = 1,\ldots, p$, then
\begin{align*}
    \Pi\Bigg( \sigma^2_{ji} \in \Big[\frac{\sigma^2_{0ji}}{1+\delta_n^2} ,\sigma^2_{0ji}\Big], \,\, j=1,\ldots,K, i = 1,\ldots, p \Bigg) & \geq \prod_{j=1}^K\prod_{i=1}^p m'_{ji} \delta_n^2 \sigma^2_{oji} \geq C_{14}\exp\{-C_{13}Kp\log(1/\delta_n) \}.
\end{align*}
Let $m''_j$ be the lower bound of the joint prior density $(\lambda_{j1},\ldots,\lambda_{jq})$ in the region of $[\lambda_{0j1},\lambda_{0ji}(1+\delta_n^4/q)]\times\cdots\times[\lambda_{0jq},\lambda_{0jq}(1+\delta_n^4/q)]$, then
\begin{align*}
    \Pi\Bigg( 1 \leq \frac{\lambda_{jk}}{\lambda_{0jk}} \leq 1+\frac{\delta_n^4}{q}, \,\, j=1,\ldots,K, k = 1,\ldots, q \Bigg) & \geq \prod_{j=1}^K m''_j \prod_{k=1}^q\lambda_{0jk}\delta_n^4/q \geq C_{16}\exp\{-C_{15}Kq\log(1/\delta_n) \}.
\end{align*}
Therefore, for sufficiently large $n$, we have
\begin{align*}
     \Pi(B_n(p_{\Theta_0},\epsilon_n))\geq \Pi(\Theta\in S_{\Theta_0})\cdot\Pi(G_j,j=1,\ldots,K) \geq C_{18}\exp\{-C_{17}K(p+q)\log(1/\delta_n) \} \geq \exp(-Cn\epsilon_n^2).
\end{align*}
\end{proof}

The conclusion of Theorem \ref{th:contraction} follows directly from Theorem 3.1 of \citet{bhattacharya2019bayesian}.

\begin{proposition}\label{normaltv} {\normalfont (Total variation distance between two normal densities \citep{devroye2018total}).}
$$ TV(N(\mu_1,\Sigma_1), N(\mu_2,\Sigma_2))\leq \frac{1}{2}\sqrt{\mbox{tr}(\Sigma^{-1}_1\Sigma_2-I) + (\mu_1-\mu_2)^T\Sigma_1^{-1}(\mu_1-\mu_2) - \log|\Sigma^{-1}_1\Sigma_2| }.$$
\end{proposition}

\begin{proposition}\label{KLbounds} {\normalfont (Upper bounding KL distance with Hellinger distance \citep{shen2013adaptive}).} There is a $\lambda_0\in(0,1)$ such that for any $\lambda\in(0,\lambda_0)$ and any two probability measures $P$ and $Q$ with respective densities $p$ and $q$,
\begin{align*}
    P\log\frac{p}{q} & \leq d^2_H(p,q)\Bigg(1+2\log\frac{1}{\lambda} \Bigg) + 2P\Bigg\{\Bigg(\log\frac{p}{q}\Bigg)\mathbbm{1}\Bigg(\frac{q}{p}\leq\lambda\Bigg)\Bigg\}, \\
    P\Bigg(\log\frac{p}{q}\Bigg)^2 & \leq d^2_H(p,q)\Bigg\{ 12 + 2\Bigg(\log\frac{1}{\lambda}\Bigg)^2\Bigg\} + 8P\Bigg\{\Bigg(\log\frac{p}{q}\Bigg)^2\mathbbm{1}\Bigg(\frac{q}{p}\leq\lambda\Bigg)\Bigg\},
\end{align*}
where $d_H(\cdot,\cdot)$ is the Hellinger distance.
\end{proposition}


\section{Posterior Inference} \label{sec:algo}
In this section, we develop a blocked Gibbs sampler for the PxG model. Local graph updating procedures are introduced for both approaches in \S \ref{sec:likelihood}. 
We detail the procedures for posterior summary and graph prediction. We also provide the criterion for model comparison between the PxG model and its nested models.

\subsection{Blocked Gibbs sampler}
As stated in \S \ref{partitionprior}, with the current choice of cohesion function $c(\cdot)$, the DP mixture model is a special case of the PxG model.
Therefore, we adopt the blocked Gibbs sampler \citep{ishwaran2001gibbs} for the PxG model and exploit the (partially) parallel computing for efficient posterior inference. 
According to \citet{sethuraman1994constructive}, sampling from a DP prior can be constructed using ``stick-breaking'' weights $V_j\sim\mbox{Beta}(1, \alpha)$, $j=1,\ldots, K-1$ and $V_K=1$ where $K$ is the maximum number of clusters allowed. The weights are then given by $\pi_1=V_1$ and
$\pi_j = V_j \prod_{l=1}^{j-1}(1-V_l)$, $j = 2,\ldots,K$. The proposed blocked Gibbs sampler consists of five steps at each iteration. 
\begin{enumerate}
\item Update the posterior stick-breaking weights $V_j$ with 
        \begin{equation*}
            V_j\sim \text{Beta}\bigg(1+n_j, \alpha + \sum_{l=j+1}^{K}n_l\bigg), \quad j=1,\ldots, K-1, \text{ and } V_K=1.
        \end{equation*} 
\item Update posterior cluster probabilities $\pi_j$ with 
        \begin{equation*}
            \pi_j = V_j \prod_{l=1}^{j-1}(1-V_l), j = 2,\ldots,K, \text{ and } \pi_1 = V_1.
        \end{equation*}
\item Update cluster indicator $z_i$ for  $i=1,\ldots,n$. Let $p_{ij}$ be the full conditional probability of assigning the $i$th sample to cluster $j$. Then we draw $z_i$ from $\{1,\ldots,K\}$ with probability $\{p_{i1},\ldots,p_{iK}\}$, where
        \begin{equation*}
            p_{ij} = P(z_i=j\mid \cdot) \propto \pi_jL(\mathrm{y}_i\mid \Omega_j, G_j) p(\mathrm{x}_i\mid \bm{\mu_{j}},\sigma_{j}^2),
        \end{equation*}
and $p(\mathrm{x}_i\mid\bm{\mu}_j,\sigma_j^2)$ is the auxiliary probabilistic model in \eqref{eq:covf}. Note that this step can be executed in parallel because of the conditional independencies of  $z_i$'s given the cluster probabilities $\pi_j$'s and the cluster-specific parameters. 
\item Update covariate-related parameters $\bm{\mu}_{j}$ and $\sigma^2_{j}$, $j=1,\ldots, K,$ from
        \begin{align*}
            \bm{\mu}_{j} \mid \sigma^2_{j} &\sim N_p(\bm{\mu}_{j}^*, \sigma^{*2}_{j} I_p), \\
            \sigma^2_{j} &\sim \mbox{IG}(b_{1j}^*, b_{2j}^*), 
        \end{align*}
where
        \begin{align*}
            \bm{\mu}_{j}^* &=\frac{\sigma_{0}^2}{n_j\sigma^2_{0}+1}\bigg(\sum_{i\in S_j}\mathrm{x}_i+\frac{\bm{\mu}_{0}}{\sigma^2_{0}}\bigg), \quad 
            \sigma^{*2}_{0} = \frac{\sigma_{0}^2\sigma^2_{j}}{n_j\sigma^2_{0}+1}, \\
            b^*_{1j} &= \frac{n_jp}{2} + b_{1}, \quad b^*_{2j} = b_{2} + \frac{1}{2}\bigg\{\sum_{i\in S_j}\mathrm{x}_i^T\mathrm{x}_i + \frac{\bm{\mu}_{0}^T\bm{\mu}_{0}}{\sigma^2_0} - \bigg(\sum_{i\in S_j}\mathrm{x}_i+\frac{\bm{\mu}_{0}}{\sigma^2_{0}}\bigg)^T\bm{\mu}^*_{j} \bigg\}.
        \end{align*}
When a cluster is empty, draw $\bm{\mu}_{j}$ and $\sigma^2_{j}$ from their priors in (\ref{eq:covf}).
\item Update cluster-specific parameters $\Omega_j$ and $G_j$ for $j=1,\ldots, K$. This local update depends on the choice of likelihood functions, see the next two subsections for detailed procedures.
\end{enumerate}
\subsubsection{Local update for the Gaussian-G-Wishart approach}
Given samples in the $j$th cluster, its cluster-specific graphs and precision matrices are updated iteratively for all $1\leq s<t\leq q$ in the following two steps.
\begin{itemize}
    \item Draw the edge inclusion indicator $g_{st}^j$ from its full conditional distribution,
        \begin{align*}
        P(g_{st}^j=a\mid \cdot)\propto p(\Omega_j \mid G_j^{st,a}) p(G_j^{st,a}) \mbox{~~for~~}a=0,1,
        \end{align*}
where $G_j^{st,a}$ is $G_j$ with $g_{st}^j=a$.
    \item Draw the precision matrix $\Omega_j$ from its full conditional distribution,
        \begin{equation*}
            \Omega_j \sim \text{G-Wishart}_{G_{j}}(b+n_j, D+{\mathrm{Y}^*_j}^T\mathrm{Y}^*_j).
        \end{equation*}
\end{itemize}

\subsubsection{Local update for the pseudo-likelihood approach}
This procedure is carried out independently (in parallel) for $q$ linear regressions. Given the $j$th cluster, for the $s$th linear regression, the update contains the following three steps.
\begin{itemize}
    \item Draw the edge inclusion indicator $g^j_{st}$ from its full conditional distribution,
        \begin{align*}
            P(g_{st}^j=a\mid \cdot)\propto N(\beta^j_{st}\mid 0, \eta_a\tau_{s}^j) p(G_j^{st,a}) \mbox{~~for~~}a=0,1 \mbox{~~and~~} t\in[q]\backslash s.
        \end{align*}
    \item Draw the residual variance $\tau_s^j \mid \bm{\beta}_s^j \sim \mbox{IG}(a_{1js}, a_{2js})$, 
    where
    \begin{align*}
        a_{1js} &= a_1 + \frac{n_j}{2} + \frac{q-1}{2}, \\
        a_{2js} &= a_2 + \frac{1}{2}(\mathrm{Y}^*_{j,s}-\mathrm{Y}^*_{j,-s}\bm{\beta}^j_s )^T(\mathrm{Y}^*_{j,s}-\mathrm{Y}^*_{j,-s}\bm{\beta}^j_s) + \frac{1}{2}(\bm{\beta}^j_s)^T A^j_{-s} \bm{\beta}^j_s,
    \end{align*}
    and $\mathrm{Y}^*_{j,s}$ is the $s$th column of $\mathrm{Y}^*_j$; $\mathrm{Y}^*_{j,-s}$ is $\mathrm{Y}^*_j$ without its $s$th column; $A^j_{-s}$ is a diagonal matrix with diagonal entries $1/\eta_{g^j_{st}}$, $t\in [q]\backslash s$.
    \item Draw the regression coefficient $\bm{\beta}^j_s\mid \tau^j_s, A^j_{-s} \sim N(\bm{m}^j_s, \bm{v}_{s}^j)$, where 
    \begin{align*}
        \bm{m}^j_s & = \big\{(\mathrm{Y}^*_{j,-s})^T\mathrm{Y}^*_{j,-s} + A^j_{-s}\big\}^{-1}(\mathrm{Y}^*_{j,-s})^T\mathrm{Y}^*_{j,s}, \\
        \bm{v}^j_s & = \tau^j_s\big\{(\mathrm{Y}^*_{j,-s})^T\mathrm{Y}^*_{j,-s} + A^j_{-s}\big\}^{-1}.
    \end{align*}
\end{itemize}

\subsubsection{Posterior graph estimation with pseudo-likelihood approach}\label{fix}
For efficient computation, we do not constrain $g_{st}^j=g_{ts}^j$ during the course of the Gibbs sampler for the pseudo-likelihood approach. As a result, $g_{st}^j$ and $g_{ts}^j$ are not necessarily the same. To restore the symmetry, we use the common approach that takes either the union or the intersection of $g_{st}^j$ and $g_{ts}^j$, i.e., $g_{st}^j=g_{ts}^j:=g_{st}^j\vee g_{ts}^j$ or $g_{st}^j=g_{ts}^j:=g_{st}^j\wedge g_{ts}^j$.

\subsection{Posterior summary and prediction}
The proposed model allows two ways of posterior inference. A clustering result can be obtained directly from the posterior samples of clustering labels. We follow the procedure in \citet{dahl2006model} which minimizes a squared loss function \citep{binder1978bayesian}. 
The cluster-specific graphs are then inferred conditionally on the estimated partition. 


Alternatively, one can also compute the graph estimate for each observation, i.e., $\widehat{G}(\mathrm{x}_i)$ by the proposed partition averaging technique in \S \ref{pagp} which is approximated by posterior samples. Specifically, let $\{\widehat{g}_{st}^{\mathrm{x}_i}\}_{s,t=1}^q$ denote the edge inclusion indicators of $\widehat{G}(\mathrm{x}_i)$ and let $\{\widehat{p}_{st}^{\mathrm{x}_i}\}_{s,t=1}^q$ denote the edge inclusion probabilities where $\widehat{p}_{st}^{\mathrm{x}_i}=P(\widehat{g}_{st}^{\mathrm{x}_i}=1)$. Then edge inclusion probability is computed by averaging the posterior samples,
\begin{equation*}
\widehat{p}_{st}^{\mathrm{x}_i}=\frac{1}{N}\sum_{r=1}^N (g_{st}^{z_i^r})^r,
\end{equation*}
where $N$ is the number of posterior samples, $z_i^r$ is the $r$th posterior sample of the cluster indicator $z_i$ and $(g_{st}^{z_i^r})^r$ is the $r$th posterior sample of the edge inclusion indicator in cluster $z_i^r$. The posterior point estimate of $\widehat{G}(\mathrm{x}_i)$ is obtained by $\widehat{g}_{st}^{\mathrm{x}_i}=I(\widehat{p}_{st}^{\mathrm{x}_i}>c)$ for some cutoff $c\in (0,1)$. For simplicity, we use $c=0.5$ (i.e., the median probability model, \citet{barbieri2004optimal}) but if desired, $c$ can be also chosen by controlling the posterior expected false discovery rate \citep{muller2006fdr}. 
When the data are generated with a definitive clustering structure, such continuous estimates of graphs reflect the estimation uncertainties due to sampling noise. Likewise, the precision matrix function $\widehat{\Omega}(\mathrm{x}_i)$ is estimated by $\widehat{\Omega}_{st}(\mathrm{x}_i)=\frac{1}{N}\sum_{r=1}^N (\Omega_{st}^{z_i^r})^r$.

For a new covariate $\mathrm{x}_\text{new}$, we can predict its graph $\widehat{G}(\mathrm{x}_{\text{new}})$ and precision matrix $\widehat{\Omega}(\mathrm{x}_{\text{new}})$ by approximating the posterior predictive distribution in \eqref{eqn:gp},
\begin{equation*}
    \widehat{p}_{st}^{\mathrm{x}_\text{new}}=\frac{1}{N}\sum_{r=1}^N (g_{st}^{z_\text{new}^r})^r, \qquad
    \widehat{g}_{st}^{\mathrm{x}_\text{new}}=I(\widehat{p}_{st}^{\mathrm{x}_\text{new}}>c), \qquad
    \widehat{\Omega}_{st}(\mathrm{x}_\text{new})=\frac{1}{N}\sum_{r=1}^N (\Omega_{st}^{z_\text{new}^r})^r.
\end{equation*}



\subsection{Model comparison}
In order to demonstrate the necessity of incorporating covariates into GGMs with heterogeneous data, we compare the proposed PxG model with two simpler models: (1) partition-based graphical model without covariates, termed \textit{graph-only model}; and (2) random partition model of covariates without the primary variables $\mathrm{y}$, termed \textit{covariate-only model}. 
Since they are both nested within the PxG model, we compare them using the deviance information criterion (DIC) \citep{spiegelhalter2002bayesian, gelman2013bayesian}. DIC accounts for both model fitting and model complexity, and the model with the smallest DIC value is deemed to be the optimum. See the next section for the formulas of all three DICs. 

\subsection{DIC for model comparison} \label{sec:DIC}
\noindent In this section, we give the DIC formula for the PxG model and its two nested model used in model comparison. \\

\noindent \textbf{DIC for the PxG model (a.k.a. full model)}
\small
\begin{equation*}
  \text{DIC}_\text{full}(\hat{\bm{\rho}},\hat{G}_1,\dots,\hat{G}_K) = -2\sum_{j=1}^{\hat{K}} \big(\log p(\mathrm{Y}^*_j\mid\hat{G}_j) + \log g(\mathrm{X}^*_j)\big) + \text{var}\Big( \sum_{j=1}^{\hat{K}} \big(\log p(\mathrm{Y}^*_j\mid\hat{G}_j)+ \log g(\mathrm{X}^*_j)\Big),
\end{equation*}
where
\begin{align*}
  \text{var}\Big( \sum_{j=1}^{\hat{K}} \big(\log p(\mathrm{Y}^*_j\mid\hat{G}_j)+ \log g(\mathrm{X}^*_j)\Big) & = 
  \frac{1}{B-1}\sum_{l=1}^B\Bigg( \sum_{j=1}^{\hat{K}_l} \big(\log p(\mathrm{Y}^*_j\mid\hat{G}^l_j)+ \log g(\mathrm{X}^*_j)\big) - U_0\Bigg)^2, \\
  U_0 & = \frac{1}{B} \sum_{l=1}^B \sum_{j=1}^{\hat{K}_l} \big(\log p(\mathrm{Y}^*_j\mid\hat{G}^l_j)+ \log g(\mathrm{X}^*_j)\big).
\end{align*}
$B$ is the number of MCMC samples; $\hat{K}_l$ is the number of clusters in the $l$th MCMC sample; $\hat{G}^l_j$ is the posterior graph sample of the $j$th cluster in the $l$th MCMC sample; $\hat{G}_j$ is the final graph estimate for the $j$th cluster given cluster allocation $\hat{\bm{\rho}}$. \\

\noindent \textbf{DIC for the graph only model (without considering covariates)}
\begin{equation*}
  \text{DIC}_G(\hat{\bm{\rho}},\hat{G}_1,\dots,\hat{G}_K) = -2\sum_{j=1}^{\hat{K}} \log p(\mathrm{Y}^*_j\mid\hat{G}_j) + \text{var}\Big( \sum_{j=1}^{\hat{K}} \log p(\mathrm{Y}^*_j\mid\hat{G}_j)\Big) - 2\log g(\mathrm{X}),
\end{equation*}
where
\begin{align*}
  \text{var}\Big( \sum_{j=1}^{\hat{K}} \log p(\mathrm{Y}^*_j\mid\hat{G}_j)\Big) & = 
  \frac{1}{B-1}\sum_{l=1}^B\Bigg( \sum_{j=1}^{\hat{K}_l} \log p(\mathrm{Y}^*_j\mid\hat{G}^l_j)- U_1\Bigg)^2, \\
  U_1 & = \frac{1}{B} \sum_{l=1}^B \sum_{j=1}^{\hat{K}_l} \log p(\mathrm{Y}^*_j\mid\hat{G}^l_j)\big).
\end{align*}

\noindent \textbf{DIC for the covariate only model (only clustering covariates)}
\begin{equation*}
  \text{DIC}_\text{cov}(\hat{\bm{\rho}},\hat{G}) = -2\sum_{j=1}^{\hat{K}} \log g(\mathrm{X}^*_j)+ \text{var}\Big( \sum_{j=1}^{\hat{K}} \log g(\mathrm{X}^*_j)\Big) -2\log p(\mathrm{Y}\mid \hat{G}) + \text{var}(\log p(\mathrm{Y}\mid \hat{G})),
\end{equation*}
where
\begin{align*}
  \text{var}\Big( \sum_{j=1}^{\hat{K}} \log g(\mathrm{X}^*_j)\Big) & = 
  \frac{1}{B-1}\sum_{l=1}^B\Bigg( \sum_{j=1}^{\hat{K}_l} \log g(\mathrm{X}^*_j)- U_2\Bigg)^2, \\
  U_2 & = \frac{1}{B} \sum_{l=1}^B \sum_{j=1}^{\hat{K}_l} \big(\log \log g(\mathrm{X}^*_j)\big). \\
  \text{var}(\log p(\mathrm{Y}\mid \hat{G})) &= \frac{1}{B-1}\sum_{l=1}^B\Bigg(\log p(\mathrm{Y}\mid \tilde{G}_l) - U'_2 \Bigg)^2, \\
  U'_2 & = \frac{1}{B} \sum_{l=1}^B \log p(\mathrm{Y}\mid \tilde{G}_l).
\end{align*}
For the covariate only model, we assume the graph is the same for all samples. Thus, we perform graph estimation with all samples $\mathrm{Y}$ and produce the same amount of MCMC samples; $\tilde{G}_l$ is the posterior graph sample in the $l$th MCMC sample.
\normalfont

\bibliographystyle{apalike}
\bibliography{PxG}

\end{document}